\documentclass{dcds}
  \usepackage{amsmath}
  \usepackage{paralist}
  \usepackage{graphics} 
  \usepackage{epsfig} 

  \usepackage{hyperref}

\def\currentvolume{}
 \def\currentissue{}
  \def\currentyear{}
   \def\currentmonth{}
    \def\ppages{}

\newtheorem{theorem}{Theorem}[section]

\newtheorem{lemma}[theorem]{Lemma}

\theoremstyle{definition}

\title[Thresholds for DNLS breathers]
      {Thresholds for breather solutions of the Discrete
  Nonlinear Schr\"odinger Equation with saturable and power
  nonlinearity}

\author[J. Cuevas, J. C. Eilbeck and N. I. Karachalios]{}

\subjclass{Primary: 37L60, 35Q55.  Secondary: 47J30.}
 \keywords{Discrete Nonlinear
    Schr\"odinger Equation, lattice differential equations, breather
    solutions, saturable nonlinearity, variational methods}

 \email{cuevas@us.es}
 \email{J.C.Eilbeck@hw.ac.uk}
\email{karan@aegean.gr}

\begin{document}
\maketitle

\centerline{\scshape J. Cuevas}
\medskip
{\footnotesize
 \centerline{Departamento de Fisica Aplicada I, Escuela Universitaria
    Polit\'{e}nica}
    \centerline{C/ Virgen de Africa, 7, University of Sevilla,}
   \centerline{University of Sevilla, 41011 Sevilla, Spain.}
} 

\medskip

\centerline{\scshape J. C. Eilbeck}
\medskip
{\footnotesize
 \centerline{Maxwell Institute and Department of Mathematics,}
 \centerline{Heriot-Watt University,}
   \centerline{Edinburgh EH14 4AS, Scotland.}}

\medskip
\centerline{\scshape N. I. Karachalios}
\medskip
{\footnotesize
 \centerline{Department of Mathematics,}
   \centerline{University of the Aegean,}
   \centerline{Karlovassi GR 83200, Samos, Greece.}}

\medskip

 \centerline{(Communicated by )}

\medskip
\begin{abstract}
We consider the question of existence of periodic solutions (called
  breather solutions or discrete solitons) for the Discrete Nonlinear
  Schr\"odinger Equation with saturable and power nonlinearity.
  Theoretical and numerical results are proved concerning the
  existence and nonexistence of periodic solutions by a variational
  approach and a fixed point argument. In the variational approach we
  are restricted to DNLS lattices with Dirichlet boundary conditions.
  It is proved that there exists parameters (frequency or nonlinearity
  parameters) for which the corresponding minimizers satisfy explicit
  upper and lower bounds on the power. The numerical studies performed
  indicate that these bounds behave as thresholds for the existence of
  periodic solutions.  The fixed point method considers the case of
  infinite lattices. Through this method, the existence of a threshold
  is proved in the case of saturable nonlinearity and an explicit
  theoretical estimate which is independent on the dimension is given.
  The numerical studies, testing the efficiency of the bounds derived
  by both methods, demonstrate that these thresholds are quite sharp
  estimates of a threshold value on the power needed for the the
  existence of a breather solution.  This it justified by the
  consideration of limiting cases with respect to the size of the
  nonlinearity parameters and nonlinearity exponents.

\end{abstract}
\section{Introduction}
This work concerns the Discrete Nonlinear Schr\"odinger Equation  (DNLS)
\begin{eqnarray}
\label{DNLSsn}
\mathrm{i}\dot{\psi}_n+\epsilon(\Delta_d\psi)_n-\beta
F(|\psi_n|^2)\psi_n =0,\;\;\beta\in\mathbb{R},
\end{eqnarray}
on a finite lattice supplemented with Dirichlet boundary conditions,
and on infinite lattices ($n=(n_1,n_2,\ldots,n_N)\in\mathbb{Z}^N$).
We concentrate on two examples of nonlinearities,
\begin{eqnarray}
\label{nonls}
F(|z|^2)=\frac{1}{1+|z|^2}\;\;\mbox{and}\;\;F(|z|^2)=|z|^{2\sigma},
\end{eqnarray}
the {\em saturable} and {\em power nonlinearity} respectively.

Note that we use the word {\em power} in two different senses in this
paper, in {\em power nonlinearity} as above, and for a conserved
quantity of the system (\ref{DNLSsn}), defined as 
\begin{equation}
\mathcal{P}[\phi]= \sum_n |\phi_n|^2. \label{power}
\end{equation}

We present some theoretical and numerical results related to the
existence of time periodic solutions, having the form
\begin{eqnarray}
\label{sw}
\psi_n(t)=e^{-\mathrm{i}\Omega t}\phi_n,\;\;\Omega\in\mathbb{R}.
\end{eqnarray}
Substitution of the expression (\ref{sw}) into (\ref{DNLSsn}) with the
nonlinearities (\ref{nonls}), shows that
$\phi=\{\phi_n\}_{n\in\mathbb{Z}^N}$, satisfies the system of
algebraic equations
\begin{eqnarray}
\label{EuLag}
\Omega\phi_n=-\epsilon(\Delta_d\phi)_n+\beta F(|\phi_n|^2)\phi_n, \quad
\beta\in\mathbb{R}.
\end{eqnarray}
Solutions given by the expression (\ref{sw}), are called stationary
wave solutions. Localized solutions fulfilling $|\phi_n|\rightarrow0$
as $|n|\rightarrow\infty$, are known as discrete solitons or
breathers\footnote{We wish to reserve the term
  ``breathers'' throughout the text, for localized solutions of the
  form (\ref{sw}), although this term is not strictly valid in the case of a
  finite lattice: this is due to the fact that the lower bounds for the power of solutions (\ref{sw})
  derived in this work in the case of finite lattice,
  are also valid for breather solutions.}. The problem of existence
and properties of nonlinear localized modes in DNLS lattices, has
attracted considerable research interest \cite{Eil, Kevrekidis}. For
recent studies on the saturable DNLS or its cubic-quintic
approximation, we refer to
\cite{EilSatDNLS,Maluckov1,Maluckov2,Maluckov3,RCarretal,Jes2,RodMagnus}.
In these references, as well as in \cite{DoriE,Mich} for the continuum
models, remarkable properties and differences between models with
power nonlinearities are reported. Although the case of the
fundamental localized solutions assumes that $\phi_n$ is real
\cite{RodMagnus}, the results that we present here consider the
existence and nonexistence of nontrivial breather solutions where
$\phi_n$ is in general complex. The existence of nontrivial breather
solutions for DNLS (\ref{DNLSsn}) will be established by variational
methods. More precisely, we apply direct variational methods \cite{CJ}
to appropriate constrained minimization problems. This approach has
been used to the focusing $N$-dimensional DNLS equation with a power
nonlinearity
\begin{eqnarray}
\label{DNLSs}
\mathrm{i}\dot{\psi}_n+\epsilon(\Delta_d\psi)_n+
\beta|\psi_n|^{2\sigma}\psi_n=0,\quad \beta>0,
\end{eqnarray}
in infinite lattices \cite{Wein99}. The results of \cite{Wein99} not
only establish the existence of nontrivial breather solutions, but in
addition the existence of a global minimum -- an excitation threshold
-- in one of the fundamental conserved energy quantities, the power
(or norm).  This minimum requires the nonlinearity exponent to be
greater than or equal to a certain critical value, depending on the
lattice dimension.  More precisely, it is proved in \cite[Theorem 3.1,
pg. 678]{Wein99}, that if $0<\sigma<\frac{2}{N}$, spatially localized
solutions (\ref{sw}) with $\Omega<0$ of arbitrary small power exist,
while if $\sigma\geq\frac{2}{N}$, there exists a ground state
excitation threshold $\mathcal{P}_{\mathrm{thresh}}$.  The result of
\cite{Wein99}, resolved the conjecture for ground state breathers of
\cite{FlachMac}. The second conserved energy quantity associated with
(\ref{DNLSs}), is the Hamiltonian
$$
\mathcal{H}_{\sigma}[\phi]=\epsilon(-\Delta_d\phi,\phi)_2-
\frac{\beta}{\sigma+1}\sum_{n\in\mathbb{Z}^N}|\phi_n|^{2\sigma+2},\;\;\beta>0.
$$
A {\em ground state} is a minimizer of the variational problem
$$
\inf\left\{\mathcal{H}_{\sigma}[\phi]\;\;:\mathcal{P}[\phi]=R^2\right\},
$$
where $(\cdot,\cdot)_2$ stands for the $\ell^2$-scalar product. The
existence of the excitation threshold was proved with the help of a
delicate discrete interpolation inequality similar to the
Gagliardo-Nirenberg inequality of the continuous case. It is proved in
\cite{Wein99} that the inequality
\begin{eqnarray}
\label{WGN1}
\sum_{n\in\mathbb{Z}^N}|\phi_n|^{2\sigma+2}\leq
C\left(\sum_{n\in\mathbb{Z}^N}| \phi_n|^2\right)^{\sigma}(-\Delta_d\phi,\phi)_2,
\end{eqnarray}
holds for $\sigma\geq\frac{2}{N}$. The excitation threshold is related
to the best constant of (\ref{WGN1}). The ground state solution has
frequency $\omega^*$ and power $\mathcal{P}_{\mathrm{thresh}}$.

After the preliminary results of Section 2, in Section 3 we consider
the DNLS equation with {\em saturable nonlinearity}. Following the
results mentioned above, for the DNLS equation with power nonlinearity
(\ref{DNLSs}), by the application of the variational approach to the
saturable DNLS, we derive both the existence of nontrivial breather
solutions, as well some bounds on the power of the minimizers. The
variational study considers the saturable DNLS, {\em supplemented with
  Dirichlet boundary conditions}. Although this is a simpler case in
comparison with the infinite lattice (where one has to deal with the
lack of compactness \cite{Wein99}), this case is of importance
especially for numerical simulations.  Since the infinite lattice
cannot be modelled numerically, numerical investigations normally
consider finite lattices with Dirichlet or periodic boundary
conditions. We note that the choice of boundary conditions only
matters, if a localized pulse is moving and collides with a boundary.

In our study of the DNLS equation with saturable nonlinearity, we
distinguish between defocusing ($\beta<0$) and focusing ($\beta>0$)
nonlinearity. For the defocusing case (Section 3.1), we consider two
variants of minimization problems: seeking for nontrivial breather
solutions $\phi_n(t)=e^{\mathrm{i}\omega t}\phi_n$, of prescribed
frequency $\omega>0$, in the first variant we consider a
minimization problem for the energy functional
$$
\mathcal{E}_{\omega}[\phi]:=\epsilon(-\Delta_d\phi,\phi)_2+
\omega\sum_{|n|\leq K}|\phi_n|^2,
$$
subject to a constraint on the logarithmic part of the saturable
Hamiltonian of the DNLS (\ref{DNLSsn}), defined by
$$
\mathcal{H}[\phi]=\epsilon(-\Delta_d\phi,\phi)_2+
\beta\sum_{|n|\leq K}\log(1+|\phi_n|^2).
$$
This proves the existence of a nontrivial minimizer $\hat{\phi}$ of
$\mathcal{E}_{\omega}$ (linear energy) and the existence of $\beta<0$
as a Lagrange multiplier, satisfying $-\beta>\omega>0$ such that
$\psi_n(t)=e^{\mathrm{i}\omega t}\hat{\phi}_n$ is a solution of the
saturable DNLS (\ref{DNLSsn}) with this $\beta$ as a nonlinearity
parameter.  We note that, {\em in contrast} to the DNLS with power
nonlinearity (\ref{DNLSs}), the frequency of the breather is limited
by the condition $\Lambda:=-\beta>\omega$, due to the resonance with
linear modes. Also in contrast with the power nonlinearity case, the
parameter $\beta$ cannot be scaled out. Due to this fact, the result
is of interest, justifying the minimization of the linear energy, and
the existence of a parameter $\beta$ for which this minimum is
attained.

The second variant for the defocusing case considers, for given
$\beta=-\Lambda<0$, the constrained minimization problem for the
Hamiltonian
$$
\inf\left\{\mathcal{H}[\phi]\;\;:\mathcal{P}[\phi]=R^2\right\},
$$
that is, we study the existence of the nontrivial breather solution as
a ground state.  This approach proves the existence of a nontrivial
minimizer $\phi^*$, at least in the parameter regime
$\Lambda>2\epsilon N$, and the existence of a frequency $\omega>0$ (as
a Lagrange multiplier), satisfying $\Lambda>\omega$, such that
$\psi_n(t)=e^{\mathrm{i}\omega t}\phi^*_n$ is a stationary wave
solution.  Moreover, it is proved in this parameter regime that there
exist frequencies, such that the corresponding nontrivial breather
solution satisfy an upper bound for the total power, depending on the
parameters $\Lambda, \epsilon, N$.  The first numerical study
performed on this parameter regime, for the behaviour of breather power
\footnote{We consider, in all the numerical studies throughout the
  paper, the power of single-site breathers (also known as
  Sievers--Takeno modes\cite{st88}), which are the breathers with the
  smallest power}, justifies the existence of a range of frequencies,
for which the upper bound of the power of the corresponding breather
solution is satisfied.

Section 3.2, is devoted to the focusing saturable nonlinearity
$\beta>0$. For this case, the existence of a nontrivial breather
solution $\psi_n(t)=e^{-i\Omega t}\tilde{\phi}_n$, is proved similarly
to Section 3.1, by considering the constrained minimization problem
for the Hamiltonian (a ground state).  Through the application of the
variational method, a simple relation involving the frequency $\Omega,
\beta$ and the power $\mathcal{P}[\tilde{\phi}]=R^2$ is derived, which
in terms of $\mathcal{P}$, provides a local lower bound on the power
of the minimizer. The result actually states that there exists
$\Omega>0$ satisfying this lower bound. We would like at this point to
distinguish between the term excitation threshold in the sense of
\cite{Wein99} and lower bounds, used throughout the text: The
variational methods used here, establish the existence of minimizers
of the energy functionals and the existence of parameters ($\beta$ or
$\Omega$) as Lagrange multipliers which are associated with these
minimizers. Furthermore they provide theoretically some {\em local}
(in the sense that they depend on $\Omega$ and $\beta$) lower bounds
on the power of the minimizers, with respect to the variation of these
parameters and not the existence of excitation thresholds on the power
in the sense of \cite{Wein99}.  To this end, the numerical studies
performed, investigate the efficiency of the lower bounds, as well
their possible behaviour as the parameters $\Omega$ and $\beta$ vary.
As a result, the second numerical study performed for the lower bound
on the power of the focusing saturable case, verifies that this bound
is actually the {\em smallest value of the power} below which we
should not expect the existence of a nontrivial breather solution for
arbitrary given $\omega$ and $\beta$ and dimension $N$. We call such
smallest values {\em thresholds} on the existence of periodic
solutions.

In the case of the focusing saturable nonlinearity $\beta>0$, the
numerical studies verify that for some parameter values, the bound
predicts the trend of the behaviour of the numerically computed power.
Moreover the numerical study in 2D-lattices, shows that the breather
solution of the focusing saturable DNLS demonstrates a similar
behaviour to that of the focusing DNLS with power nonlinearity, with
respect to the existence of excitation thresholds: power decreases as
the frequency increases until it reaches a minimum value at a certain
frequency, {\em an excitation threshold in the sense of
  \cite{Wein99}}. This behaviour should occur in higher dimensional
lattices, and is observed in \cite{RodMagnus}. It is worth taking into
account that the saturable nonlinearity can be approximated by a power
one with $\sigma=1$ for small values of $|\phi_n|$. This relation
holds when $\Omega$ is close to $\beta$ in the saturable case, and can
justify the similarities between the saturable and power
nonlinearities related to thresholds.

In Section 4, we apply an alternative method to derive a threshold on
the power of the breather solution of the saturable DNLS in the
focusing case $\beta>0$.  We use a fixed point argument which was also
used in \cite{K1}.  This approach is for the saturable DNLS,
considered in {\em infinite lattices} (although similar estimates can
be obtained in the case of Dirichlet boundary conditions). Replacing
the saturable nonlinearity in the equation by its exact Taylor
polynomial of order $m$, it is possible to derive the threshold for
the power for both the DNLS with the saturable nonlinearity and the
cubic-quintic approximation. The threshold appears to be the positive
root of a polynomial equation.  In the case of the 1D-lattice, the
numerical study verifies first that the power decreases as frequency
increases, as it was predicted by the lower bound derived by the
variational method of Section 3. Also, the numerical power approaches
the predicted threshold derived by the fixed point argument as the
frequency increases up to the limit $\beta$. In comparison with the
bound derived by the variational method, the latter also reaches the
threshold derived by the fixed point argument, as the frequency
increases.  Especially for large values of the parameter $\beta$, the
theoretical estimates are proved to be quite sharp for large values of
frequencies.  For 2D-lattices we also observe numerically the
appearance of the excitation threshold in the sense of \cite{Wein99}.
In the focusing case, it follows from the numerical study that an
increase of the dimension as well as of the nonlinearity parameter is
required for this excitation threshold to appear.

Section 5 is devoted to some theoretical and numerical results,
related to the DNLS equation with {\em power nonlinearity}.  We
consider first the case of the defocusing ($\beta>0$) DNLS, seeking
for breather solutions $\psi_n(t)=e^{-\mathrm{i}\Omega t}$,
$\Omega>0$. By applying the same variational approach as for the
saturable DNLS, we derive lower bounds on the power of the minimizers,
depending on the dimension of the lattice. The numerical study
demonstrates that these lower bounds can serve actually as thresholds
on the existence of breather solutions. The approach of minimizing the
linear energy functional
$$
\mathcal{E}_{\Omega}[\phi]:=\epsilon(-\Delta_d\phi,\phi)_2
-\Omega\sum_{|n|\leq K}|\phi_n|^2,\;\;\Omega>0,
$$
appears to be useful also in the power case as the numerical study
indicates, since the derived lower bound on the power gives a result
slightly closer to the real power.

We would like to summarize by pointing out the main differences with
the results of \cite{Wein99}.  The results of \cite{Wein99} prove the
existence of an excitation threshold on the power of periodic
solutions of the focusing DNLS with power nonlinearity which appears
for the case $\sigma>2/N$, as well the existence of a frequency
$\omega^*>0$ on which this threshold value on the power is achieved.
The corresponding solution $\psi_n(t)=e^{i\omega^*t}\phi_n$ is a
ground state having power $\mathcal{P}_{\mathrm{thresh}}$ -- the
excitation threshold value.

The thresholds proved by the fixed-point argument are {\em explicitly
  given threshold values which are independent of the dimension}.
There are explicit estimates satisfied for {\em any} periodic solution
with frequency $\beta>\Omega>0$ and for any $N\geq 1$ in the focusing
saturable nonlinearity, and for any $\omega>0$ in the case of the
focusing power nonlinearity and for any $\sigma>0, N\geq 1$. They are
thresholds in the sense that {\em no periodic localized solution can
  have power less than the prescribed estimates}. A characteristic
example for the justification of this claim as well as its usefulness,
is provided by the numerical study regarding the focusing power
nonlinearity. {\em Even in the case $\sigma<2/N$, where the excitation
  threshold \cite{Wein99} does not exist}, a periodic solution cannot
have power less than the derived estimate. This ``global character''
of the estimates is revealed when one considers ``limiting'' cases of
small (large) values of $\sigma<2/N$ (large values of $\sigma\geq
2/N$ -- the case of excitation threshold). The numerical studies verify
that for small (large) values of frequencies the threshold is not only
satisfied but is also a quite sharp estimate of the real power of the
corresponding periodic solutions. Especially in the case $\sigma>2/N$,
the numerical studies demonstrate the fact that the excitation
threshold $\mathcal{P}_{\mathrm{thresh}}$, which is not known
explicitly, satisfies the derived lower bound. Thus this bound should
not be viewed as a prediction of the excitation threshold in the case
of $\sigma\geq 2/N$, nor as a theoretical prediction of the numerical
power of periodic solutions, but as prediction of the smallest power a
periodic solution can have, for any $\omega,\sigma$ and $N\geq 1$. The
same global property is shared by the estimate for the saturable
nonlinearity, as the numerical study in the case of large values of
$\beta$ in the defocusing case shows. We conclude with the remark that
the numerical studies for the focusing saturable nonlinearity, as well
as for the defocusing power nonlinearity, show  that the local
estimates derived by the variational methods also predict the smallest
value of the power for arbitrary given values of parameters
$\beta,\sigma,\omega$ and $N$.

\section{Preliminaries}
For the convenience of the reader, we recall
some basic results on sequence spaces and their finite dimensional
subspaces as well on discrete operators, that will be used in the
sequel (see also \cite{K1,Wein99}).

For some positive integer $N$, we consider the complex sequence spaces
\begin{equation}
\label{ususeqs}
\ell^p=\left\{
\begin{array}{ll}
&\phi=\{\phi_n\}_{n\in\mathbb{Z}^{N}},\;n=(n_1,n_2,\ldots,n_N)
   \in\mathbb{Z}^N,\;\;\phi_n\in\mathbb{C},\\
&||\phi||_{p}=\left(\sum_{n\in\mathbb{Z}^N}|\phi_n|^p\right)^{\frac{1}{p}}
   <\infty
\end{array}
\right\}.
\end{equation}
(The following elementary embedding relation \cite{ree79} holds between
$\ell^p$ spaces
\begin{eqnarray}
\label{lp1}
\ell^q\subset\ell^p,\;\;\;\; ||\phi||_{p}\leq ||\phi||_{q}\,\;\;
1\leq q\leq p\leq\infty,
\end{eqnarray}
in contrast with the $L^p(\Omega)$-spaces, if
$\Omega\subset\mathbb{R}^N$ has finite measure. For $p=2$, we get the
usual Hilbert space of square-summable sequences, which becomes a real
Hilbert space if endowed with the real scalar product
\begin{eqnarray}
\label{lp2}
(\phi,\psi)_{2}=\mathrm{Re}\sum_{{n\in\mathbb{Z}^N}}\phi_n
   \overline{\psi_n},\;\;\phi,\,\psi\in\ell^2.
\end{eqnarray}
Note that any $\phi_n\in\ell^p$, $1\leq p<\infty$ satisfies
$\lim_{|n|\rightarrow\infty}\phi_n=0$, as assumed for spatially
localized solutions (i.e. discrete solitons or breathers).
The discrete Laplacian is defined for $\phi_n=\phi_{(n_1,n_2,\ldots,n_N)}$ as
\begin{eqnarray}
\label{DiscLap}
(\Delta_d\psi)_{n\in\mathbb{Z}^N}&=&\psi_{(n_{1}-1,n_2,\ldots ,n_N)}
  +\psi_{(n_1,n_{2}-1,\ldots ,n_N)}+\cdots+
\psi_{(n_1,n_{2},\cdots ,n_N-1)}\nonumber\\
&&-2N\psi_{(n_{1},n_2,\ldots ,n_N)}
+\psi_{(n_{1}+1,n_2,\ldots ,n_N)}\nonumber\\
&&+\psi_{(n_1,n_{2}+1,\ldots ,n_N)}+\cdots+
\psi_{(n_1,n_{2},\cdots ,n_N+1)},
\end{eqnarray}
Now we consider the discrete operator $\nabla^+:
\ell^2\rightarrow\ell^2$ defined by
\begin{eqnarray}
\label{discder1}
(\nabla^+\psi)_{n\in\mathbb{Z}^N}&=&\left\{\psi_{(n_{1}+1,n_2,\ldots
    ,n_N)}
  -\psi_{(n_{1},n_2,\ldots ,n_N)}\right\}\nonumber\\
&+&\left\{\psi_{(n_{1},n_2+1,\ldots ,n_N)}
  -\psi_{(n_{1},n_2,\ldots ,n_N)}\right\}\nonumber\\
&\vdots&\nonumber\\
&+&\left\{\psi_{(n_{1},n_2,\ldots ,n_N+1)}
  -\psi_{(n_{1},n_2,\ldots ,n_N)}\right\},
\end{eqnarray}
and $\nabla^{-}:\ell^2\rightarrow\ell^2$ defined by
\begin{eqnarray}
\label{discder2}
(\nabla^-\psi)_{n\in\mathbb{Z}^N}&=
&\left\{\psi_{(n_{1}-1,n_2,\ldots ,n_N)}
  -\psi_{(n_{1},n_2,\ldots ,n_N)}\right\}\nonumber\\
&+&\left\{\psi_{(n_{1},n_2-1,\ldots ,n_N)}
  -\psi_{(n_{1},n_2,\ldots ,n_N)}\right\}\nonumber\\
&\vdots&\nonumber\\
&+&\left\{\psi_{(n_{1},n_2,\ldots ,n_N-1)}
  -\psi_{(n_{1},n_2,\ldots ,n_N)}\right\}.
\end{eqnarray}
Setting
\begin{eqnarray}
\label{discder3}
(\nabla^+_{\nu}\psi)_{n\in\mathbb{Z}^N}=\psi_{(n_{1},n_2,\ldots ,
n_{\nu-1},n_{\nu}+1,n_{\nu+1},\ldots ,n_N)}-\psi_{(n_{1},n_2,\ldots ,n_N)},\\
\label{discder4}
(\nabla^-_{\nu}\psi)_{n\in\mathbb{Z}^N}=\psi_{(n_{1},n_2,\ldots ,
n_{\nu-1},n_{\nu}-1,n_{\nu+1},\ldots ,n_N)}-\psi_{(n_{1},n_2,\ldots ,n_N)},
\end{eqnarray}
we observe that the operator $-\Delta_d$ satisfies the relations
\begin{eqnarray}
\label{diffop2}
(-\Delta_d\psi_1,\psi_2)_{2}&=&\sum_{\nu=1}^N(\nabla_\nu^+\psi_1,\nabla_\nu^+
\psi_2)_{2},\;\;\mbox{for all}\;\;\psi_1,\psi_2\in\ell^2,\\
\label{diffop3}
(\nabla_\nu^+\psi_1,\psi_2)_{2}&=&(\psi_1,\nabla_\nu^-\psi_2)_{2},\;\;
\mbox{for all}\;\;\psi_1,\psi_2\in\ell^2.
\end{eqnarray}
{}From (\ref{diffop2}), it is clear that $-\Delta_d: \ell^2
\rightarrow\ell^2$ defines a self adjoint operator on $\ell^2$,
and $-\Delta_d\geq 0$.

To formulate the DNLS equation, subject to Dirichlet boundary
conditions, we consider the finite dimensional subspaces of $\ell^p$
for a positive integer $K$, defined by
\begin{eqnarray}
\label{subs}
\ell^p(\mathbb{Z}^N_K)=\left\{\phi\in\ell^p\;:\;\phi_n=0\;\;
\mbox{for}\;\;|n|>K\right\}.
\end{eqnarray}
We have $\ell^p(\mathbb{Z}^N_K)\equiv \mathbb{C}^{(2K+1)^N}$ endowed with the norms
(\ref{lp1})-finite sums.  In this case, for any $1\leq p\leq
q\leq\infty$, there exist constants $C_1,C_2$ depending on $K$, such
that
\begin{eqnarray}
\label{fnorms}
C_1||\psi||_{p}\leq ||\psi||_{q}\leq C_2||\psi||_{p}.
\end{eqnarray}
In the finite dimensional setting, the operator $-\Delta_d$ satisfies
relations (\ref{diffop2})-(\ref{diffop3}), and its principal eigenvalue
$\lambda_1>0$ can be characterized as
\begin{eqnarray}
\label{eigchar}
\lambda_1=\inf_{
\begin{array}{c}
\phi \in \ell^2(\mathbb{Z}^N_K) \\
\phi \neq 0
\end{array}}\frac{(-\Delta_d\phi,\phi)_{2}}{(\phi,\phi)_{2}} =
\inf_{
\begin{array}{c}
\phi \in \ell^2(\mathbb{Z}^N_K) \\
\phi \neq 0
\end{array}}
\frac{\sum_{\nu=1}^{N}||\nabla^+_{\nu}\phi||_{2}^2}{\sum_{|n|\leq K}|\phi_n|^2}.
\end{eqnarray}
Hence (\ref{eigchar}) and (\ref{discder1})-(\ref{discder2}) imply
the inequality
\begin{eqnarray}
\label{crucequiv}
\epsilon\lambda_1\sum_{|n|\leq K}|\phi_n|^2\leq
\epsilon\sum_{\nu=1}^{N}||\nabla^+_{\nu}\phi||_{2}^2\leq
4\epsilon N \sum_{|n|\leq K}|\phi_n|^2.
\end{eqnarray}
Then it follows from (\ref{crucequiv}) that
\begin{eqnarray}
\label{prineig}
\lambda_1\leq 4N.
\end{eqnarray}
For example, in the case of an $1D$ lattice $n=1,\ldots, K$, the
eigenvalues of the discrete Dirichlet problem
$-\Delta_d\phi=\lambda\phi$ with $\phi$ real, are given by
$$
\lambda_n=4\sin^2\left(\frac{n\pi}{4(K+1)}\right),\;\;n=1,\ldots,K.
$$
For a N-dimensional problem, the eigenvalues are:
\begin{eqnarray*}
&&\lambda_{(n_{1},n_{2},\ldots,n_{N})}\\
&&=4\left[\sin^2
\left(\frac{n_1\pi}{4(K+1)}\right)+\sin^2\left(\frac{n_2\pi}
{4(K+1)}\right)+\ldots+\sin^2\left(\frac{n_N\pi}{4(K+1)}\right)\right],
\end{eqnarray*}
for $n_j=1,\ldots,K,\,j=1,\ldots,N$. In consequence, the principal eigenvalue of the discrete Dirichlet
problem $-\Delta_d\phi=\lambda\phi$ with $\phi$ real, is given by
\begin{eqnarray*}
\lambda_1\equiv\lambda_{(1,1,\ldots,1)}=
4N\sin^2\left(\frac{\pi}{4(K+1)}\right).
\end{eqnarray*}

\section{Saturable nonlinearity: Constrained minimization
  problems-Dirichlet boundary conditions}
\subsection{A. Defocusing case $\beta <0$: Periodic solutions
  $\psi_n(t)=e^{i\omega t}\phi_n, \omega>0$}
In this section we consider the existence of breather solutions of the
saturable DNLS equation, for the case $\beta<0$.  For convenience we
set $$\beta=-\Lambda, \Lambda>0.$$ Thus, we seek breather solutions for the
DNLS equation
\begin{eqnarray}
\label{DNLSL}
\mathrm{i}\dot{\psi}_n+\epsilon(\Delta_d\psi)_n+
\Lambda\frac{\psi_n}{1+|\psi_n|^2}=0,\;\;\Lambda>0,
\end{eqnarray}
of the form
\begin{eqnarray}
\label{caseL}
\psi_n(t)=e^{i\omega t}\phi_n, \omega>0.
\end{eqnarray}
In this case, the system (\ref{EuLag}) is rewritten as
\begin{eqnarray}
\label{om1}
-\epsilon(\Delta_d\phi)_n+\omega\phi_n=
 \Lambda\frac{\phi_n}{1+|\phi_n|^2},\;\;\Lambda>0.
\end{eqnarray}
We note that in the case of the anticontinuum limit
$\epsilon=0$, it follows that the frequency of a non-trivial breather
solution should satisfy
\begin{eqnarray}
\label{SOS}
\Lambda>\omega.
\end{eqnarray}

The Hamiltonian
$\mathcal{H}$ and the power $\mathcal{P}$, given by
\begin{eqnarray}
\label{H1}
\mathcal{H}[\phi]&=&\epsilon\sum_{\nu=1}^{N}||\nabla^+_{\nu}\phi||_{2}^2
-\Lambda\sum_{|n|\leq K}\log(1+|\phi_n|^2),\\
\label{P}
\mathcal{P}[\phi]&=&\sum_{|n|\leq K}|\phi_n|^2,
\end{eqnarray}
are quantities which are independent of time.  We shall prove the
existence of nontrivial breather solutions (\ref{caseL}), by considering a
constrained minimization problem.  The system (\ref{om1}) will be
considered as the Euler-Lagrange equation of the functional
\begin{eqnarray}
\label{Hamil2}
\mathcal{H}_{\omega}[\phi]=
\epsilon\sum_{\nu=1}^{N}||\nabla^+_{\nu}\phi||_{2}^2
+{\omega}\sum_{|n|\leq K}|\phi_n|^2
-\Lambda\sum_{|n|\leq K}\log(1+|\phi_n|^2),
\end{eqnarray}
involving $\mathcal{H}$ and $\mathcal{P}$.  To produce the
Euler-Lagrange equation (\ref{om1}) from the functionals $\mathcal{H}$
and $\mathcal{P}$, we shall use the following
\begin{lemma}
\label{gatdevl}
Let $\phi\in\ell^{2}(\mathbb{Z}^N_K)$. Then the functional
$$
\mathcal{V}(\phi)=\sum_{|n|\leq K}\log(1+|\phi_n|^{2}),
$$
is a $\mathrm{C}^{1}(\ell^{2}(\mathbb{Z}^N_K),\mathbb{R})$ functional and
\begin{eqnarray}
\label{gatdev}
\langle\mathcal{V}'(\phi),\psi\rangle=2\mathrm{Re}\sum_{|n|\leq K}
\frac{\phi_n}{1+|\phi_n|^2}\overline{\psi_n},\;\;
\psi\in\ell^{2}(\mathbb{Z}^N_K).
\end{eqnarray}
\end{lemma}
{\bf Proof:}\ \ We assume that $\phi,\,\psi\in\ell^{2}(\mathbb{Z}^N_K)$.
Then for any $0<s<1$, we get
\begin{eqnarray}
\label{mv}
&&\frac{\mathcal{V}(\phi +s\psi)-\mathcal{V}(\phi)}{s}=
\frac{1}{s}\mathrm{Re}\sum_{|n|\leq K}\int_{0}^{1}
\frac{d}{d\theta}\log(1+|\phi_n +
\theta s\psi_n|^{2})d\theta\\
&&\;\;\;\;\;\;\;\;\;\;\;\;\;\;\;\;\;\;\;\;\;\;=
2\mathrm{Re}\sum_{|n|\leq K}\int_{0}^{1}\frac{\phi_n+s\theta\psi_n}
{1+|\phi_n+s\theta\psi_n|^2}
\overline{\psi_n} d\theta.\nonumber
\end{eqnarray}
{}From the inequality
\begin{eqnarray}
\label{mv1}
\sum_{|n|\leq K}\frac{|\phi_n+\theta s\psi_n|}{1+|\phi_n
+\theta s\psi_n|^2}|\psi_n|
\leq
\sum_{|n|\leq K}\left(|\phi_n|+|\psi_n|\right)|\psi_n|\leq (||\phi||_{\ell^2}
+||\psi||_2)||\psi||_2,
\end{eqnarray}
we may let $s\rightarrow 0$, to get the existence of the Gateaux
derivative (\ref{gatdev}) (discrete dominated convergence).

To check that the functional
$\mathcal{V}':\ell^{2}\rightarrow\ell^{2}$ is continuous, we consider
a sequence $\phi_m\in\ell^2(\mathbb{Z}^N_K)$ such that
$\phi_m\rightarrow \phi$ in $\ell^2(\mathbb{Z}^N_K)$. Then, by using
an inequality very similar to (\ref{Taylor14}) (see Section
\ref{ness}), we may show that
$\left|\left<\mathcal{V}'(\phi_m)-\mathcal{V}'(\phi),
    \,\psi\right>\right|\rightarrow 0$, as $m\rightarrow\infty$. The
result is also valid  in the case of an infinite lattice
($n\in\mathbb{Z}^N$).\ \ $\diamond$
\paragraph{\textbf{The constrained minimization problem A.I}}
The first variational problem we shall discuss is a constrained
minimization problem for the energy quantity
\begin{eqnarray}
\label{ln0}
\mathcal{E}_{\omega}[\phi]:=\epsilon\sum_{\nu=1}^{N}||\nabla^+_{\nu}
\phi||_{2}^2+\omega\sum_{|n|\leq K}|\phi_n|^2,\;\;\phi\in\ell^2
(\mathbb{Z}^N_K),\;\;\omega>0,
\end{eqnarray}
for given $\omega>0$. We have the following
\setcounter{theorem}{1}
\begin{theorem}
\label{exL}
Let $\omega>0$ be given.
Consider the variational problem on $\ell^2(\mathbb{Z}^N_K)$
\begin{eqnarray}
\label{inflognorm}
\inf\left\{\mathcal{E}_{\omega}[\phi]\;:\sum_{|n|\leq K}
\log(1+|\phi_n|^2)=R>0\right\}.
\end{eqnarray}
There exists a minimizer $\hat{\phi}\in\ell^2(\mathbb{Z}^N_K)$ for the
variational problem (\ref{inflognorm}) and a $\Lambda(R)>0$ such that
$\Lambda(R)>\omega$, both satisfying the Euler-Lagrange equation
\begin{eqnarray*}
\label{EL1a}
-\epsilon(\Delta_d\hat{\phi})_n+\omega\hat{\phi}_n&=
  &\Lambda\frac{\hat{\phi_n}}{1+|\hat{\phi}_n|^2},\;\; |n|\leq K,\\
\hat{\phi}_n&=&0,\;\;|n|>K.
\end{eqnarray*}
Moreover, it holds that $\sum_{|n|\leq K}\log(1+|\phi_n|^2)=R$.
\end{theorem} {\bf Proof:}\ Clearly $\mathcal{E}_{\omega}$ defines a
$C^1(\ell^2(\mathbb{Z}^N_K),\mathbb{R})$ functional. The minimization
problem we intend to solve is for the functional
$\mathcal{E}_{\omega}$, restricted to the set
\begin{eqnarray}
\label{ln1}
B_1=\left\{\phi\in\ell^2(\mathbb{Z}^N_K)\;:\;
\sum_{|n|\leq K}\log(1+|\phi_n|^2)=R>0\right\},\;\;
\end{eqnarray}
It is not hard to check that the sequence
$\{\phi_m\}_{m\in\mathbb{N}}\in B_1$ is bounded. Next, we consider a
sequence $\{\phi_m\}_{m\in\mathbb{N}}\in B_1$, such that
$\phi_m\rightarrow\phi$ as $m\rightarrow\infty$. We denote the $n$th
coordinate of this sequence by $(\phi_m)_n$. Using (\ref{fnorms}), we
observe that
\begin{eqnarray}
\label{ln2}
\left|\sum_{|n|\leq K}\log(1+|(\phi_m)_n|^2)-\sum_{|n|\leq K}
\log(1+|\phi_n|^2)\right|\leq
C||\phi_m-\phi||_{\ell^1(\mathbb{Z}^N_K)}
\rightarrow 0,
\end{eqnarray}
as $m\rightarrow\infty$. Moreover, since $\phi_m\in B_1$, we find from (\ref{ln2}) that
\begin{eqnarray*}
\sum_{|n|\leq K}\log(1+|\phi_n|^2)=\lim_{m\rightarrow\infty}
\sum_{|n|\leq K}\log(1+|(\phi_m)_n|^2)=R.
\end{eqnarray*}
Hence $\phi\in B_1$, which implies that $B_1$ is closed.  The
functional $\mathcal{E}_{\omega}$ is bounded from below on
$\mathcal{B}_1$, since
\begin{eqnarray}
\label{bbl1}
\mathcal{E}_{\omega}[\phi]&=&\epsilon\sum_{\nu=1}^{N}||\nabla^+_{\nu}
\phi||_{2}^2+\omega\sum_{|n|\leq K}|\phi_n|^2\nonumber\\
&&\geq\omega\sum_{|n|\leq K}|\phi_n|^2\nonumber\\
&&\geq \omega \sum_{|n|\leq K}\log(1+|\phi_n|^2)\geq \omega R.
\end{eqnarray}
As we are restricted to the finite dimensional space
$\ell^2(\mathbb{Z}^N_K)$, it follows that any minimizing sequence
associated with the variational problem (\ref{inflognorm}) is
precompact: any minimizing sequence has a subsequence, converging to a
minimizer.  Thus $\mathcal{E}_{\omega}$ attains its infimum at a point
$\hat{\phi}$ in $B_1$. We proceed in order to derive the variational
equation for $\mathcal{E}_{\omega}$. Note that
\begin{eqnarray}
\label{ln3}
\left<\mathcal{E}_{\omega}'[\phi],\psi\right>=
2\epsilon\sum_{\nu=1}^N(\nabla_\nu^+\phi,\nabla_\nu^+\psi)_{2}
+2\omega\mathrm{Re}\sum_{|n|\leq K}\phi_n\overline{\psi}_n.
\end{eqnarray}
By considering the $C^1(\ell^2(\mathbb{Z}^N_K),\mathbb{R})$
(see Lemma \ref{gatdev})
\begin{eqnarray}
\label{ln4}
\mathcal{L}_R[\phi]=\sum_{|n|\leq K}\log(1+|\phi_n|^2)-R,
\end{eqnarray}
we observe that for any $\phi\in B_1$
\begin{eqnarray}
\label{ln5}
\left<\mathcal{L}_R'[\phi],\phi\right>=2\sum_{|n|\leq
  K}\frac{|\phi_n|^2}{1+|\phi_n|^2}>0.
\end{eqnarray}
The Regular Value Theorem (\cite[Section 2.9]{CJ}, \cite[Appendix
A,pg. 556 ]{speight}) implies that the set $B_1=\mathcal{L}_R^{-1}(0)$
is a $C^{1}$-submanifold of $\ell^2(\mathbb{Z}^N_K)$. By applying the
Lagrange multiplier rule, we find the existence of a parameter
$\Lambda=\Lambda(R)\in\mathbb{R}$, such that
\begin{eqnarray}
\label{ln6}
\left<\mathcal{E}_{\omega}'[\hat{\phi}]-\Lambda\mathcal{L}_R'[\hat{\phi}],
\psi\right>&=&
2\epsilon\sum_{\nu=1}^N(\nabla_\nu^+\hat{\phi},\nabla_\nu^+\psi)_{2}
+2\omega\mathrm{Re}\sum_{|n|\leq K}\hat{\phi}_n\overline{\psi}_n\nonumber\\
&&-2\Lambda\mathrm{Re}\sum_{|n|\leq K}\frac{\hat{\phi}_n}
{1+|\hat{\phi}_n|^2}\overline{\psi}_n=0,
\end{eqnarray}
for all
$\psi\in\ell^2(\mathbb{Z}^N_K)$. Setting $\psi=\hat{\phi}$ in (\ref{ln6}), we find that
\begin{eqnarray}
\label{ln7}
2\mathcal{E}_{\omega}[\hat{\phi}]=2\epsilon\sum_{\nu=1}^{N}||\nabla^+_{\nu}
\hat{\phi}||_{2}^2+2\omega\sum_{|n|\leq K}|\hat{\phi}_n|^2
=2\Lambda\sum_{|n|\leq K}\frac{|\hat{\phi}_n|^2}{1+|\hat{\phi}_n|^2}.
\end{eqnarray}
Since $\hat{\phi}\in B_1$ cannot be identically zero and
$\mathcal{E}_{\omega}[\hat{\phi}]>0$, it follows from (\ref{ln7}) that
$\Lambda>0$.

Rewriting (\ref{ln7}) as
\begin{eqnarray}
\label{rest7e}
\epsilon\sum_{\nu=1}^{N}||\nabla^+_{\nu}\hat{\phi}||_{2}^2
=\Lambda\sum_{|n|\leq K}\frac{|\hat{\phi}_n|^2}{1+|\hat{\phi}_n|^2}-
\omega\sum_{|n|\leq K}|\hat{\phi}_n|^2,
\end{eqnarray}
we find that
\begin{eqnarray}
\label{7f}
\epsilon\sum_{\nu=1}^{N}||\nabla^+_{\nu}\hat{\phi}||_{2}^2\leq
(\Lambda-\omega)\sum_{|n|\leq K}|\hat{\phi}_n|^2.
\end{eqnarray}
The lhs of (\ref{7f}) is positive, and we find that
\begin{eqnarray}
\Lambda(R)>\omega.
\end{eqnarray}
Therefore condition (\ref{SOS}) is justified.  From (\ref{ln6}),
there exists $\Lambda>0$, such that the minimizer $\hat{\phi}\in B_1$
solves the equation
\begin{eqnarray}
\label{weakS2}
\left(-\epsilon\Delta_{\mathrm{d}}\phi,\psi\right)_{2}+\omega(\phi,\psi)_{2}=
\Lambda\left(\frac{\phi}{1+|\phi|^2},\psi\right)_{2},\;\;\mbox{for
  all}\;\;
\psi\in\ell^2(\mathbb{Z}^N_K).
\end{eqnarray}
The above formula, is clearly equivalent to the Euler-Lagrange
equation (\ref{om1}), that is, any solution of (\ref{om1}) is a
solution of (\ref{weakS2}), and vice versa.\ \ $\diamond$
\paragraph{\textbf{The constrained minimization problem A.II}}
In the first minimization problem, we derived, under sufficient
conditions for given $\omega>0$, both the existence of
$\beta=-\Lambda<0$ as a Lagrange multiplier, and the existence of a
nontrivial $\hat{\phi}\in\ell^2(\mathbb{Z}^N_K)$, such that
(\ref{caseL}) is a breather solution of (\ref{DNLSL}). Here, by studying a
different minimization problem, we shall derive some sufficient
conditions depending on the given $\beta=-\Lambda<0$, the lattice
spacing $\epsilon$ and the dimension of the lattice $N$.  These provide
both the existence of a parameter $\omega>0$ and a nontrivial
$\phi^*$, such that (\ref{swO}) is a solution of (\ref{DNLSsn})
involving this $\omega$ as the frequency of the breather solution.
This alternative variational approach for the existence of breather
solutions (\ref{caseL}) for the DNLS equation (\ref{DNLSL}), is to
minimize the Hamiltonian $\mathcal{H}$, constrained to the set
\begin{eqnarray}
\label{restri1} B=\left\{\phi\in\ell^2(\mathbb{Z}^N_K)\;:\;
\sum_{|n|\leq K}|\phi_n|^2=R^2>0\right\}.
\end{eqnarray}
\begin{theorem}
\label{exomega}
Let $\Lambda,\epsilon, N>0$ be chosen such that
\begin{eqnarray}
\label{paramsA}
\Lambda>4\epsilon N.
\end{eqnarray}
Consider the variational problem on $\ell^2(\mathbb{Z}^N_K)$
\begin{eqnarray}
\label{infnorm}
\inf\left\{\mathcal{H}[\phi]\;:\mathcal{P}[\phi]=R^2>0\right\}.
\end{eqnarray}
Assuming that
\begin{eqnarray}
\label{paramsB}
R^2<\frac{\Lambda-4\epsilon N}{4\epsilon N},
\end{eqnarray}
there exists a minimizer $\phi^*\in\ell^2(\mathbb{Z}^N_K)$ for the
variational problem (\ref{infnorm}) and $\omega=\omega(R)>0$ such that
$\Lambda>\omega(R)$, both satisfying the Euler-Lagrange equation
\begin{eqnarray*}
\label{EL1}
-\epsilon(\Delta_d\phi^*)_n+\omega\phi_n^*&=&\Lambda\frac{\phi_n^*}
{1+|\phi_n^*|^2},\;\; |n|\leq K,\\
\phi^*_n&=&0,\;\;|n|>K.
\end{eqnarray*}
Moreover, it holds that $\mathcal{P}[\phi^*]=R^2$.
\end{theorem}
{\bf Proof:}
We note first that $\mathcal{H}:\ell^2(\mathbb{Z}^N_K)
\rightarrow\mathbb{R}$, is bounded from below, since
\begin{eqnarray}
\label{restri2}
\mathcal{H}[\psi]\geq -\Lambda\sum_{|n|\leq K}\log(1+|\phi_n|^2)\geq
-\Lambda\sum_{|n|\leq K}|\phi_n|^2=-\Lambda R^2.
\end{eqnarray}
Again, the finite dimensionality of the problem implies that any
minimizing sequence associated with the variational problem
(\ref{restri1}) is precompact, and that any minimizing sequence has a
subsequence, which converges to a minimizer.  Therefore, we conclude
that $\mathcal{H}:B\rightarrow\mathbb{R}$ attains its infimum at a
point $\phi^*\in B$.  The next step is to derive the variational
equation (\ref{EuLag}).  To this end, by using Lemma \ref{gatdevl}, we
observe that
\begin{eqnarray}
\label{rest4}
\left<\mathcal{H}'[\phi],\psi\right>=2\epsilon\sum_{\nu=1}^N
(\nabla_\nu^+\phi,\nabla_\nu^+\psi)_{2}-2\Lambda\mathrm{Re}
\sum_{|n|\leq K}\frac{\phi_n}{1+|\phi_n|^2}\overline{\psi}_n.
\end{eqnarray}
We consider next the functional $\mathcal{N}_R:\ell^2(\mathbb{Z}^N_K)
\rightarrow\mathbb{R}$
\begin{eqnarray}
\label{rest5}
\mathcal{N}_R[\phi]:=\sum_{|n|\leq K}|\phi_n|^2-R^2.
\end{eqnarray}
Clearly, $\mathcal{N}_R$ is $C^1(\ell^2(\mathbb{Z}^N_K),\mathbb{R})$ and
\begin{eqnarray}
\label{rest6}
\left<\mathcal{N}_R'[\phi],\psi\right>=2\mathrm{Re}\sum_{|n|\leq
  K}\phi_n \overline{\psi}_n.
\end{eqnarray}
Moreover, for any $\phi\in B$, we have
\begin{eqnarray}
\label{rest7}
\left<\mathcal{N}_R'[\phi],\phi\right>=2\sum_{|n|\leq K}|\phi_n|^2=2R^2\neq 0.
\end{eqnarray}
Therefore, we may apply the Regular Value Theorem, to show that the
set $B=\mathcal{N}_R^{-1}(0)$ is a $C^1$-submanifold of
$\ell^2(\mathbb{Z}^N_K)$.  It follows from (\ref{rest4}),
(\ref{rest6}) and the rule of Lagrange multipliers, that there exists a
Lagrange multiplier $\Omega=\Omega(R)\in \mathbb{R}$, such that
\begin{eqnarray}
\label{rest8}
\left<\mathcal{H}'[\phi^*]-\Omega\mathcal{N}_R'[\phi^*],\psi\right>&= &
2\epsilon\sum_{\nu=1}^N(\nabla_\nu^+\phi^*,\nabla_\nu^+\psi)_{2}
-2\Lambda\mathrm{Re}\sum_{|n|\leq K}\frac{\phi_n^*}{1+|\phi_n^*|^2}
\overline{\psi}_n\nonumber\\
&&-2\Omega\mathrm{Re}\sum_{|n|\leq K}\phi_n^*\overline{\psi_n}=0,\;\;
\mbox{for all}\;\;\psi\in\ell^2(\mathbb{Z}^N_K).
\end{eqnarray}
Thus, for $\psi=\phi^*$, we find that
\begin{eqnarray}
\label{rest9}
2\mathcal{U}(\phi^*)=2\Omega\sum_{|n|\leq K}|\phi^*|^2=2\Omega R^2,
\end{eqnarray}
where the functional $\mathcal{U}:\ell^2(\mathbb{Z}^N_K)
\rightarrow\mathbb{R}$ is defined by
\begin{eqnarray*}
\mathcal{U}(\phi):=\epsilon\sum_{\nu=1}^{N}||\nabla^+_{\nu}\phi||_{2}^2
-\Lambda\sum_{|n|\leq K}\frac{|\phi_n|^2}{1+|\phi_n|^2}.
\end{eqnarray*}
For an appropriate choice of $R,\epsilon,\Lambda$, we can show that
$\mathcal{U}[\phi^*]<0$: From (\ref{crucequiv}), we have
\begin{eqnarray}
\label{rest9a}
\epsilon\sum_{\nu=1}^{N}||\nabla^+_{\nu}\phi||_{2}^2\leq
4\epsilon N||\phi||_2^2.
\end{eqnarray}
Then, noting that $|\phi^*_n|^2\leq \sum_{|n|\leq K}|\phi^*_n|^2$, and
using (\ref{rest9a}), we observe that
\begin{eqnarray}
\label{rest9b}
\mathcal{U}(\phi^*)\leq 4\epsilon N\sum_{|n|\leq K}|\phi_n^*|^2-
\Lambda\sum_{|n|\leq K}\frac{|\phi_n^*|^2}{1+||\phi_n^*||_2^2}=
\left(4\epsilon NR^2-\Lambda\frac{R^2}{1+R^2}\right).
\end{eqnarray}
Therefore $\mathcal{U}[\phi^*]<0$, if
\begin{eqnarray}
\label{rest9c}
4\epsilon N(1+R^2)<\Lambda.
\end{eqnarray}
which in terms of $R$, gives (\ref{paramsB}). We shall consider
equation (\ref{rest9}), for the choice of parameters (\ref{rest9c}):
since $\phi^*\in B$ cannot be identically zero, and
$\mathcal{U}[\phi^*]<0$, it follows that $\Omega(R)<0$. Thus, we may
set
\begin{eqnarray}
\label{rest9d}
\Omega(R)=-\omega(R),\;\;\omega(R)>0.
\end{eqnarray}

The inequality (\ref{7f}) is still applicable, to verify that
$\omega(R)$ satisfies condition (\ref{SOS}). We get from
(\ref{rest8}) and (\ref{rest9d}), that there exists $\omega(R)>0$,
such that the minimizer $\phi^*\in B$ solves the equation
(\ref{weakS2}).\ \ $\diamond$

\paragraph{Numerical Study.} The result of Theorem \ref{exL},
establishes the existence of a nontrivial $\hat{\phi}$, for a given
$\omega>0$, and the existence of $\Lambda>0$, satisfying
$\Lambda>\omega$ such that $\psi_n(t)=e^{i\omega t}\hat{\phi}_n$,
solves the (\ref{DNLSL}) equation, with such a $\Lambda$ as a
nonlinearity parameter. In Theorem \ref{exL}, there are no further
relations assumed between $\omega, \epsilon, N$.

On the other hand, the result of Theorem \ref{exomega} has the
following implementation concerning the existence of breather solutions: for
given $\Lambda,\epsilon, N$ satisfying condition (\ref{paramsA}),
$$
\Lambda>4\epsilon N,
$$
{\em there exists some $\omega>0$} such that $\Lambda>\omega$ and
$\phi^*\in\ell^2(\mathbb{Z}^N_K)$ not identically zero, solving the
Euler-Lagrange equation (\ref{om1}). Hence, if (\ref{paramsA}) holds,
there exists $\omega>0$, such that $\psi_n(t)=e^{i\omega t}\phi_n^*$,
is a solution for the DNLS equation (\ref{DNLSL}), with power
$$
\mathcal{P}[\psi]=R^2<\frac{\Lambda-4\epsilon N}{4\epsilon N}.
$$
The statement of the Theorem \ref{exomega} could be useful in the
sense that the parameter regime (\ref{paramsA}) establishes the
existence of a range for frequencies $\omega>0$, such that the
corresponding breather solutions of the DNLS equation (\ref{DNLSL}) have
power satisfying the upper bound (\ref{paramsB}).

\begin{figure}
\begin{center}
    \begin{tabular}{cc}
    \includegraphics[scale=0.34]{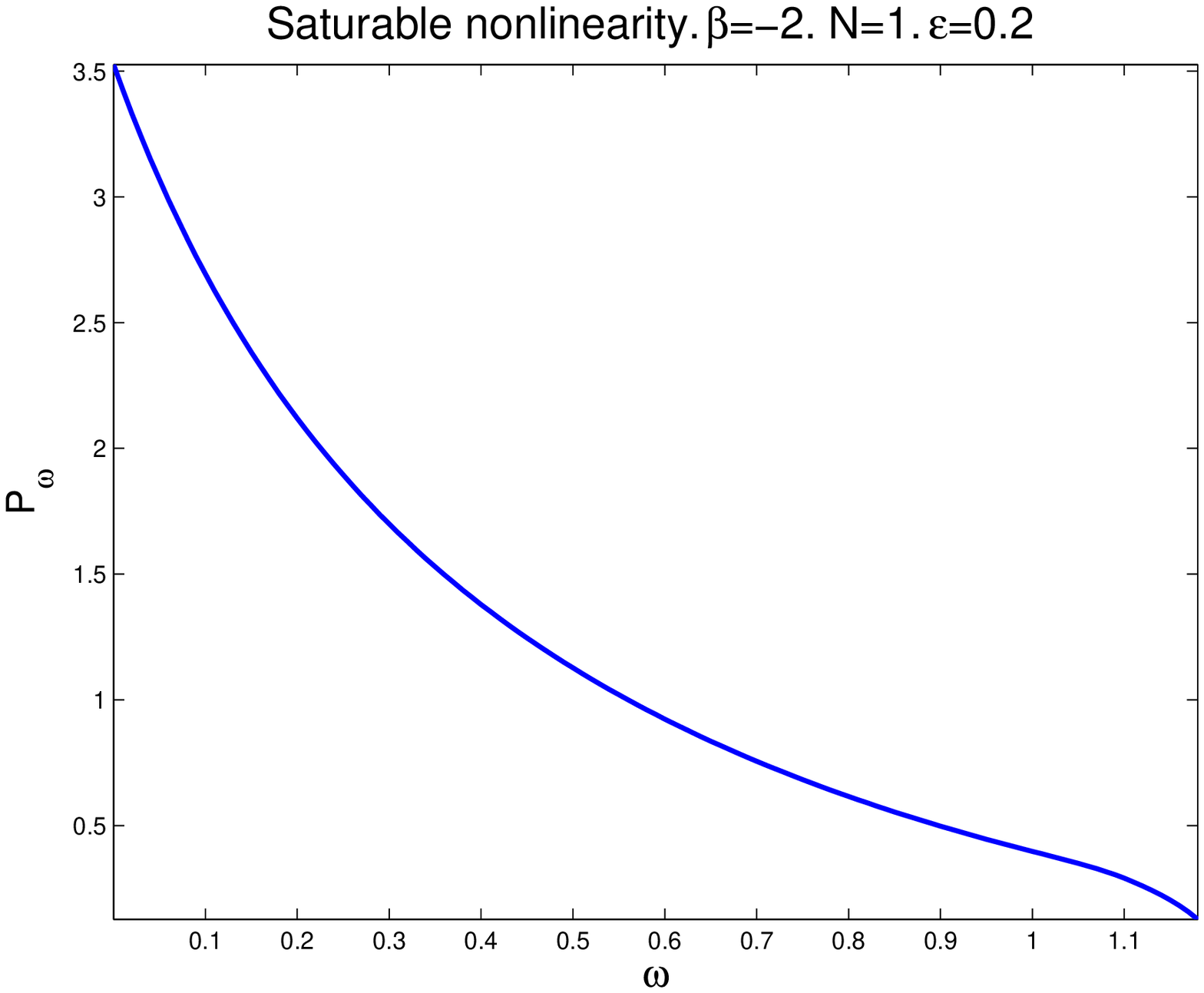} &
    \includegraphics[scale=0.34]{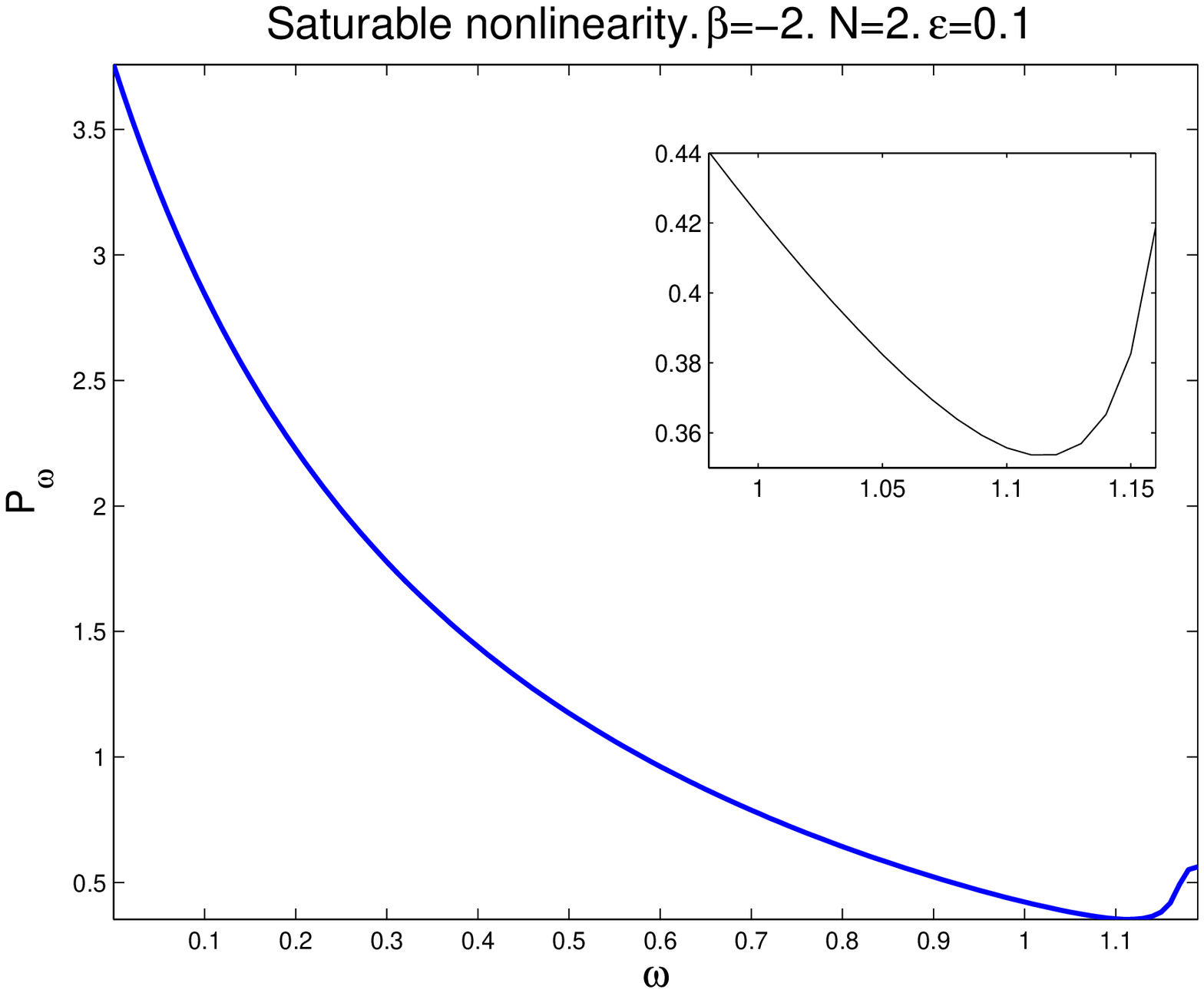}
    \end{tabular}
    \caption{Power vs frequency of the defocusing DNLS with saturable
      nonlinearity for the cases (a) $\Lambda=2,\epsilon=0.2, N=1$ (b)
      $\Lambda=2,\epsilon=0.1, N=2$.  The predicted value by the upper
      bound (\ref{paramsB}), in both cases is $\mathcal{P}=1.5$ (see
      text). In case (b), we observe the existence of a minimum of the
      total power (excitation threshold)}.
    \label{fig:upper}
\end{center}
\end{figure}

A numerical study has been performed to study the behaviour of the power of
the breather solution in the parameter regime (\ref{paramsA}), to test the
result of Theorem \ref{exomega} and the upper bound (\ref{paramsB}).
There is an extra condition for the existence of breathers arising
from the condition of non-resonance with linear modes, which is that
\begin{eqnarray}
\label{nonres}
\Lambda>4N\epsilon-\omega.
\end{eqnarray}
Clearly, (\ref{nonres}) is satisfied when (\ref{paramsA}) is
assumed.  The numerical power verifies that there exists a range of
frequencies such that the corresponding breather solutions have
power satisfying the upper bound (\ref{paramsB}):  first we have
depicted the power versus the frequency for breathers with
$\Lambda=2$ and $\epsilon=0.2$ in a 1-dimensional lattice (see Fig.
\ref{fig:upper}). From the theoretical prediction, an $\omega>0$
should exist satisfying $\Lambda>\omega$, with a power which should
be always smaller than 1.5. From the figure, it can be deduced that
the prediction is satisfied for all $\omega>0.353$. We have also
considered the case $\Lambda=2$, $\epsilon=0.1$, for $N=2$.
Similarly, for the 2D-lattice, the breathers solution of frequency
$\omega>0.830$, satisfy the theoretical upper bound
$\mathcal{P}=1.5$. The numerical study in the 2D-case reveals the
existence of an excitation threshold for the defocusing saturable
DNLS as in the case of the power nonlinearity \cite{Wein99}: power
decreases as frequency increases, attaining a minimum value for a
certain value of frequency, as shown in the inset of Figure
\ref{fig:upper} (b). As the frequency increases further, it seems
that the power reaches a ``threshold value''. In the case of higher
dimensional lattices, the upper bound (\ref{paramsB}),  could be
even more useful, as an estimate {\em from above} of the excitation
threshold as well as of the ``threshold'' value of the increased
power reached, as the frequency increases further up to the resonant
limit.
\subsection{B. Focusing case $\beta >0$: Periodic solutions
  $\psi_n(t)=e^{-i\Omega t}\phi_n, \Omega>0$}
This section considers the existence of breather solutions of the saturable
DNLS equation in the focusing case $\beta>0$.  We look for breather
solutions of the form
\begin{eqnarray}
\label{swO}
\psi_n(t)=e^{-i\Omega t}\phi_n, \quad \Omega>0
\end{eqnarray}
By considering again the case of the anticontinuum limit $\epsilon=0$,
it follows that the frequency of a nontrivial breather solution (\ref{swO})
satisfies
\begin{eqnarray}
\label{HII0}
\beta>\Omega.
\end{eqnarray}
We choose to consider the minimization problem for the Hamiltonian
\begin{eqnarray}
\label{HII1}
\mathcal{H}[\phi]&=&\epsilon\sum_{\nu=1}^{N}||\nabla^+_{\nu}\phi||_{2}^2
+\beta\sum_{|n|\leq K}\log(1+|\phi_n|^2),\;\;\beta>0,
\end{eqnarray}
constrained on the set $B$ given by (\ref{restri1}), since the
approach of the variational problem A.I does not seem to be
applicable in this case.  \setcounter{theorem}{3}
\begin{theorem}
\label{exOmega}
Let $\beta>0$ be given, and
consider the following variational problem on $\ell^2(\mathbb{Z}^N_K)$
\begin{eqnarray}
\label{HIIm1}
\inf\left\{\mathcal{H}[\phi]\;:\mathcal{P}[\phi]=R^2>0\right\}.
\end{eqnarray}
There exists a minimizer $\tilde{\phi}\in\ell^2(\mathbb{Z}^N_K)$ for
the variational problem (\ref{infnorm}) and $\Omega=\omega(R)>0$ such
that $\beta>\Omega(R)$, both satisfying the Euler-Lagrange equation
\begin{eqnarray*}
\label{HIIm2}
-\epsilon(\Delta_d\tilde{\phi})_n-\Omega\tilde{\phi}_n&=&
  -\beta\frac{\tilde{\phi}_n}{1+|\tilde{\phi}_n|^2},\;\; |n|\leq K,\\
\tilde{\phi}_n&=&0,\;\;|n|>K.
\end{eqnarray*}
Moreover, it holds that $\mathcal{P}[\tilde{\phi}]=R^2$ and
\begin{eqnarray}
\label{HIIt2}
\Omega>\frac{\beta}{1+R^2}.
\end{eqnarray}
\end{theorem} {\bf Proof:} To see that $\mathcal{H}$, is bounded from
below, this time we  use the inequality (\ref{eigchar}): we have
\begin{eqnarray}
\label{HII2}
\mathcal{H}[\phi]&\geq&\epsilon
\sum_{\nu=1}^{N}||\nabla^+_{\nu}\phi||_{2}^2 \nonumber\\
&\geq&
\epsilon\lambda_1\sum_{|n|\leq K}|\phi_n|^2\geq\epsilon\lambda_1 R^2.
\end{eqnarray}
The existence of the minimizer $\tilde{\phi}\in\ell^2(\mathbb{Z}^N_K)$,
and of the Lagrange multiplier $\Omega\in\mathbb{R}$, can be derived
by using the same variational arguments as in case A: since
\begin{eqnarray}
\label{HII3}
\left<\mathcal{H}'[\tilde{\phi}]-\Omega\mathcal{N}_R'[\tilde{\phi}],
\psi\right>&=&
2\epsilon\sum_{\nu=1}^N(\nabla_\nu^+\tilde{\phi},\nabla_\nu^+\psi)_{2}
+2\beta\mathrm{Re}\sum_{|n|\leq K}\frac{\tilde{\phi}_n}
{1+|\tilde{\phi}_n|^2}\overline{\psi}_n\nonumber\\
&&-2\Omega\mathrm{Re}\sum_{|n|\leq K}\tilde{\phi}_n\overline{\psi_n}=0,\;\;
\mbox{for all}\;\;\psi\in\ell^2(\mathbb{Z}^N_K).
\end{eqnarray}
Setting $\psi=\tilde{\phi}$, we get the equation
\begin{eqnarray}
\label{HII4}
2\mathcal{U}[\tilde{\phi}]:=2\epsilon\sum_{\nu=1}^{N}||\nabla^+_{\nu}
\tilde{\phi}||_{2}^2
+2\beta\sum_{|n|\leq K}\frac{|\tilde{\phi}_n|^2}{1+|\tilde{\phi}_n|^2}=
2\Omega\sum_{|n|\leq K}|\tilde{\phi}_n|^2,
\end{eqnarray}
and since $\tilde{\phi}\neq 0$ and
$\mathcal{U}[\tilde{\phi}]>0$, we find that $\Omega(R)>0$. To justify
condition (\ref{HII0}), we note by using (\ref{rest9a}) and
(\ref{HII4}), that for arbitrary $\epsilon>0$
\begin{eqnarray*}
(4\epsilon N+\beta)\sum_{|n|\leq K}|\tilde{\phi}_n|^2> \Omega
\sum_{|n|\leq K}|\tilde{\phi}_n|^2,
\end{eqnarray*}
implying that
\begin{eqnarray*}
\epsilon+\frac{\beta}{4N}> \frac{\Omega}{4N},\;\;\mbox{for any}
\;\;\epsilon>0,
\end{eqnarray*}
thus
\begin{eqnarray}
\label{HII0a}
\beta\geq\Omega>0.
\end{eqnarray}
Since $\mathcal{P}[\tilde{\phi}]=R^2$, we get from (\ref{HII4}), that
\begin{eqnarray}
\label{HIIt1}
2\epsilon\sum_{\nu=1}^{N}||\nabla^+_{\nu}\tilde{\phi}||_{2}^2&=&
2\Omega\sum_{|n|\leq K}|\tilde{\phi}_n|^2-2\beta\sum_{|n|\leq K}
\frac{|\tilde{\phi}_n|^2}{1+|\tilde{\phi}_n|^2}\nonumber\\
&&\leq 2\Omega\sum_{|n|\leq K}|\tilde{\phi}_n|^2-\sum_{|n|\leq K}
\frac{|\tilde{\phi}_n|^2}{1+||\tilde{\phi}_n||_2^2}
\nonumber\\
&&\leq 2R^2\left(\Omega-\frac{\beta}{1+R^2}\right).
\end{eqnarray}
Let us assume that $R>0$ and
\begin{eqnarray}
\label{atopon}
\Omega\leq\frac{\beta}{1+R^2}.
\end{eqnarray}
Then, (\ref{HIIt1}) and (\ref{crucequiv}) imply that $R=0$, and in this case (\ref{atopon}) turns out that $\beta\leq\Omega$.  Thus, we have a contradiction both with the assumption $R>0$ as well as with (\ref{HII0}).  Therefore for $R>0$, we should have $\beta>\Omega$ and (\ref{HIIt2}).\ \ $\diamond$
\vspace{0.2cm}\newline On the other hand, the inequality
(\ref{HIIt2}), in terms of the power, it can be rewritten
\begin{eqnarray}
\label{HIIt3}
R^2>\frac{\beta}{\Omega}-1.
\end{eqnarray}
Thus the result of Theorem \ref{exOmega} shows that, for given
$\beta>0$, there exists some $\Omega>0$, satisfying $\beta>\Omega$,
and a nontrivial $\tilde{\phi}\in\ell^2(\mathbb{Z}^N_K)$, such that
$\psi_n(t)=e^{-i\Omega t}\tilde{\phi}_n$, is a breather
solution of (\ref{DNLSsn}), with power satisfying the lower bound
(\ref{HIIt3}). Let us observe that the rhs of (\ref{HIIt3}) predicts
that the power should be a decreasing function of the frequency
$\Omega$, as the frequency increases to the resonant limit $\beta$.
In the next section, we shall derive a threshold value for the
power of breather solutions for the focusing case $\beta>0$, by
using a fixed point argument. The lower bound of this section as
well as the fixed point threshold, will be tested numerically.

\section{Thresholds for periodic solutions of the saturable DNLS by a
  fixed point argument-Infinite lattices: focusing case $\beta>0$}
\label{ness}
We repeat here the fixed point argument of
\cite{K1} to derive a threshold on the power, for the non-existence of
non-trivial breather solutions for (\ref{DNLSsn}). The approach covers the
case of an infinite lattice $(n\in\mathbb{Z}^N)$.  We consider the
case where the parameters $\beta>\Omega>0$ are given, and we
investigate conditions on the non-existence of non-trivial solutions
of the form
\begin{eqnarray}
\label{posfreq}
\psi_n(t)=e^{-\mathrm{i}\Omega t}\phi_n,\;\;\beta>\Omega>0.
\end{eqnarray}
Note that $\phi$ satisfies (\ref{EuLag}), rewritten as
\begin{eqnarray}
\label{EuLag2}
-\epsilon(\Delta_d\phi)_n-\Omega\phi_n=
-\beta\frac{\phi_n}{1+|\phi_n|^2},\;\;\beta>\Omega>0.
\end{eqnarray}
For the convenience of the reader we state \cite[Theorem 18.E, pg.
68]{zei85} (Theorem of Lax and Milgram), which as for the case of the
$2\sigma$-power nonlinearity \cite{K1}, will be used to establish
existence of solutions for an auxiliary linear system of algebraic
equations related to (\ref{EuLag2}).
\begin{theorem}
\label{LMth}
Let $X$ be a Hilbert space and $\mathbf{A}:X\rightarrow X$ be a linear
continuous operator. Suppose that there exists $c^*>0$ such that
\begin{eqnarray}
\label{strongmonot}
\mathrm{Re}(\mathbf{A}u,u)_X\geq c^*||u||^2_X,\;\;\mbox{for all}\;\;u\in X.
\end{eqnarray}
Then for given $f\in X$, the operator equation $\mathbf{A}u=f,\;\;u\in
X$, has a unique solution
\end{theorem}
Recall that for $f(x)=\frac{1}{1+x^2}, x\in\mathbb{R}$, the following
identity holds
\begin{eqnarray}
\label{Tiden}
f(x)=1-x^2+x^4+\cdots+(-1)^nx^{2n}+\frac{(-1)^{m+1}x^{2(m+1)}}{1+x^2},
\quad
\mbox{for all}\;\; x\in\mathbb{R},
\end{eqnarray}
which coincides with the Taylor polynomial of order $m$. Applying (\ref{Tiden}) for $x=|\zeta|,\zeta\in\mathbb{C}$
we rewrite the saturable nonlinearity as
\begin{eqnarray}
\label{Taylor1}
F(\zeta)&=&\frac{\zeta}{1+|\zeta|^2}\nonumber\\
&=&\left(1-|\zeta|^2+|\zeta|^4+
\cdots+(-1)^{m}|\zeta|^{2m}
+\frac{(-1)^{m+1}|\zeta|^{2(m+1)}}{1+|\zeta|^{2}}\right)\zeta,
\end{eqnarray}
for all $\zeta\in\mathbb{C}$. Using (\ref{Taylor1}), equation (\ref{EuLag2}) can be rewritten as
\begin{eqnarray}
\label{Taylor2}
-\epsilon(\Delta_d\phi)_n+(\beta-\Omega)\phi_n= F_{*}(\phi_n)+T_{*}(\phi_n),
\end{eqnarray}
where
\begin{eqnarray}
\label{Taylor3}
F_{*}(\phi_n)&:=&\beta\left(|\phi_n|^2\phi_n-|\phi_n|^4\phi_n+
\cdots+(-1)^{m+1}|\phi_n|^{2m}\phi_n\right),\\
\label{Taylor4}
T_{*}(\phi_n)&:=&\beta\frac{(-1)^{m+2}|\phi_n|^{2(m+1)}\phi_n}{1+|\phi_n|^2}.
\end{eqnarray}
Setting
\begin{eqnarray}
\label{Taylor4a}
\delta:=\beta-\Omega>0,
\end{eqnarray}
we observe that the (linear and continuous) operator
$\mathcal{T}_{\delta}:\ell^2\rightarrow\ell^2$, defined as
\begin{eqnarray}
\label{Taylor5}
(\mathcal{T}_{\delta}\phi)_{n\in\mathbb{Z}^N}&=&
  -\epsilon(\Delta_d\phi)_{n\in\mathbb{Z}^N}+\delta\phi_n,
\end{eqnarray}
satisfies condition (\ref{strongmonot}) if (\ref{Taylor4a}) holds, since
\begin{eqnarray}
\label{Taylor6}
(\mathcal{T}_{\delta}\phi,\phi)_{2}=\epsilon\sum_{\nu=1}^N||
{\nabla}^+_{\nu}\phi||^2_{2}+\delta ||\phi||^2_2\geq
\delta||\phi||^2_{2}\;\;\mbox{for all}\;\;\phi\in\ell^2.
\end{eqnarray}
Next, setting $F_{2m}(\zeta)=|\zeta|^{2m}\zeta$, we may define from
$F_{2m}$, a map $\mathbf{F}_{2m}:\ell^2\rightarrow\ell^2$.  We have
that
\begin{eqnarray}
\label{Taylor7}
||\mathbf{F}_{2m}(z)||^2_{\ell^2}\leq\sum_{n\in\mathbb{Z}^N}|z_n|^{4m+2}
\leq||z||_2^{4m+2}.
\end{eqnarray}
Then, writing $F_*=\beta\sum_{j=1}^{m}F_{2j}$, we may also define from
$F_*$ a nonlinear map $\mathbf{F}_*:\ell^2\rightarrow\ell^2$, since
\begin{eqnarray}
\label{Taylor8}
||\mathbf{F}_*(z)||_{\ell^2}\leq\beta\sum_{j=1}^m||
\mathbf{F}_{2j}(z)||_{\ell^2}\leq\beta\sum_{j=1}^m||z||_2^{2j+1}.
\end{eqnarray}
Similarly, from the remainder term $T_*$, we may define a map
$\mathbf{T}_*:\ell^2\rightarrow\ell^2$: we have
\begin{eqnarray}
\label{Taylor9}
||\mathbf{T}_*(\phi)||^2_{\ell^2}\leq\beta^2
\sum_{n\in\mathbb{Z}^N}|\phi_n|^{4m+4}\leq\beta^2||\phi||^{4m+4}_{\ell^2}.
\end{eqnarray}
Hence the assumptions of Theorem \ref{LMth}, are satisfied and the
auxiliary linear problem
\begin{eqnarray}
\label{Taylor10}
(\mathcal{T}_{\delta}\phi)_{n\in\mathbb{Z}^N}=\mathcal{K}(z_n),\;\;
(\mathcal{K}(z))_{n\in\mathbb{Z}^N}:=(\mathbf{F}_*(z)))_{n\in\mathbb{Z}^N}
+(\mathbf{T}_*(z))_{n\in\mathbb{Z}^N},
\end{eqnarray}
has a unique solution.

We proceed with the definition of the map $\mathcal{L}$ on which the fixed point argument will be applied.
For any given $z\in\ell^2$, we define the
map $\mathcal{L}:\ell^2\rightarrow\ell^2$, by $\mathcal{L}(z):=\phi$,
where $\phi$ is the unique solution of the operator equation
(\ref{Taylor10}). Thus, $\mathcal{L}:\ell^2\rightarrow\ell^2$ is well defined. We will verify next that  $\mathcal{L}:\mathcal{B}_R\rightarrow\mathcal{B}_R$ and is a contraction, satisfying the hypotheses of the Banach fixed point theorem.

Let $\zeta$, $\xi\in \mathcal{B}_R$ such that
$\phi=\mathcal{L}(\zeta)$, $\psi=\mathcal{L}(\xi)$. The difference
$\chi:=\phi-\psi$ satisfies the equation
\begin{eqnarray}
\label{Taylor11}
(\mathcal{T}_{\delta}\chi)_{n\in\mathbb{Z}^N}=
(\mathcal{K}(\zeta))_{n\in\mathbb{Z}^N}-(\mathcal{K}(\xi))_{n\in\mathbb{Z}^N},
\end{eqnarray}
where
\begin{eqnarray*}
(\mathcal{K}(\zeta))_{n\in\mathbb{Z}^N}-(\mathcal{K}
(\xi))_{n\in\mathbb{Z}^N}&=&\left\{(\mathbf{F}_*(\zeta))_{n\in\mathbb{Z}^N}
-(\mathbf{F}_*(\xi))_{n\in\mathbb{Z}^N}\right\}\nonumber\\
&+&\left\{(\mathbf{T}_*(\zeta))_{n\in\mathbb{Z}^N}-
(\mathbf{T}_*(\xi))_{n\in\mathbb{Z}^N}\right\}.
\end{eqnarray*}
The map $\mathbf{F}_*:\ell^2\rightarrow\ell^2$ is locally Lipschitz:
We recall that for any $F\in \mathrm{C}(\mathbb{C},\mathbb{C})$ which
takes the form $F(z)=g(|\zeta|^2)\zeta$, with $g$ real and
sufficiently smooth, the following relation holds
\begin{eqnarray}
\label{GLTaylor12}
F(\zeta)-F(\xi)=\int_{0}^{1}\left\{(\zeta-\xi)(g(r)+rg'(r))
+(\overline{\zeta}-\overline{\xi})\Phi^2 g'(r)\right\}d\theta,
\end{eqnarray}
for any $\zeta,\;\xi\in \mathbb{C}$,where $\Phi=\theta \zeta
+(1-\theta)\xi$, $\theta\in (0,1)$ and $r=|\Phi|^2$ (see \cite[pg.
202]{GiVel96}).  Applying (\ref{GLTaylor12}) for the case of
$F_{2m}(\zeta)=|\zeta|^{2m}\zeta$, one finds that
\begin{eqnarray}
\label{Taylor13}
F_{2m}(\zeta)-F_{2m}(\xi)=\int_0^1[(m
+1)(\zeta-\xi)|\Phi|^{2m}
+m(\overline{\zeta}-\overline{\xi})\Phi^2|\Phi|^{2m
-2}]d\theta.
\end{eqnarray}
Assuming that $\zeta$, $\xi\in \mathcal{B}_R$, and noting that
$||\Phi||_2\leq R$, we get from (\ref{Taylor13}) the inequality
\begin{eqnarray}
\label{Taylor14}
\sum_{n\in\mathbb{Z}^N}|\mathbf{F}_{2m}(\zeta_n)-
\mathbf{F}_{2m}(\xi_n)|^2&\leq& \beta^2(2m+1)^2\sum_{n\in\mathbb{Z}^N}
\left\{\int_0^1|\Phi_n|^{2m}|\zeta_n-\xi_n|d\theta\right\}^2\nonumber\\
&\leq&
\beta^2(2m+1)^2\sum_{n\in\mathbb{Z}^N}\left\{\int_0^1||\Phi||_2^{2m}|
\zeta_n-\xi_n|d\theta\right\}^2\nonumber\\
&\leq&
\beta^2(2m+1)^2\sum_{n\in\mathbb{Z}^N}\left\{\int_0^1 R^{2m}|\zeta_n-\xi_n|
d\theta\right\}^2\nonumber\\
&&=\beta^2(2m +1)^2R^{4m}\sum_{n\in\mathbb{Z}^N}|\zeta_n-\xi_n|^2.
\end{eqnarray}
Application of (\ref{GLTaylor12}) to the remainder term, where
$$
g(r)=\frac{(-1)^{m+2}r^{m+1}}{1+r},
$$
implies that
\begin{eqnarray}
\label{Taylor15}
&&T_*(\zeta)-T_*(\xi)=\int_{0}^{1}(\zeta-\xi)
\left[\frac{(-1)^{m+2}(m+2)|\Phi|^{2m+2}}{1+|\Phi|^2}
-\frac{(-1)^{m+2}|\Phi|^{2m+4}}{(1+|\Phi|^2)^2}\right]d\theta\nonumber\\
&&+\int_{0}^{1}(\overline{\zeta}-\overline{\xi})\Phi^2
\left[\frac{(-1)^{m+2}(m+1)|\Phi|^{2m}}{1+|\Phi|^2}
-\frac{(-1)^{m+2}|\Phi|^{2m+2}}{(1+|\Phi|^2)^2}\right]d\theta.
\end{eqnarray}
Then working similarly as for the derivation of (\ref{Taylor14}), we
get that
\begin{eqnarray}
\label{Taylor16}
&&\sum_{n\in\mathbb{Z}^N}|\mathbf{T}_{*}(\zeta_n)-\mathbf{T}_{*}(\xi_n)|^2\nonumber\\
&&\leq \beta^2\sum_{n\in\mathbb{Z}^N}\left\{\int_0^1
\left[(2m+3)|\Phi_n|^{2m+2}+2|\Phi_n|^{2m+4}\right]
|\zeta_n-\xi_n|d\theta\right\}^2\nonumber\\
&&\leq
\beta^2\sum_{n\in\mathbb{Z}^N}\left\{\int_0^1\left[(2m+3)
||\Phi||^{2m+2}_{\ell^2}+2||\Phi||_{\ell^2}^{2m+4}\right]|
\zeta_n-\xi_n|d\theta\right\}^2\nonumber\\
&&\leq
\beta^2\left[[(2m+3)R^{2m+2}+2R^{2m+4}\right]^2\sum_{n\in\mathbb{Z}^N}|
\zeta_n-\xi_n|^2.
\end{eqnarray}
{}From inequalities (\ref{Taylor14}) and (\ref{Taylor16}), we set
\begin{eqnarray}
\label{Taylor17}
L_1(R)&=&\sum_{j=1}^m(2j+1)R^{2j},\nonumber\\
L_2(R)&=&(2m+3)R^{2(m+1)}+2R^{2(m+2)},\\
L(R)&=&L_1(R)+L_2(R).\nonumber
\end{eqnarray}
Then combining (\ref{Taylor14}) and (\ref{Taylor16}), we observe that
the map $\mathcal{K}:\ell^2\rightarrow\ell^2$ is locally Lipschitz,
satisfying
\begin{eqnarray}
\label{Taylor18}
||\mathcal{K}(\zeta)-\mathcal{K}(\xi)||_{\ell^2}\leq\beta L(R)
||\zeta-\xi||_{\ell^2}.
\end{eqnarray}
Taking now the scalar product of (\ref{Taylor11}) with $\chi$ in
$\ell^2$ and using (\ref{Taylor18}), we have that
\begin{eqnarray}
\label{Taylor19}
\epsilon\sum_{\nu=1}^{N}||\nabla^+_\nu\chi||_{\ell^2}^2+\delta||\chi||^2_2&
\leq& ||\mathcal{K}(\zeta)-\mathcal{K}(\xi)||_{2}||\chi||_{2}\nonumber\\
&\leq&\beta L(R)||\zeta-\xi||_{2}||\chi||_{2}\nonumber\\
&\leq&\frac{\delta}{2}||\chi||_{2}^2+\frac{\beta^2}{2\delta}L^2(R)
||\zeta-\xi||_{2}^2.
\end{eqnarray}
{}From (\ref{Taylor19}), we obtain the inequality
\begin{eqnarray}
\label{Taylor20}
||\chi||_{2}^2=||\mathcal{L}(z)-\mathcal{L}(\xi)||_{2}^2
\leq \frac{\beta^2}{\delta^2}L^2(R)||\zeta-\xi||^2_{2}.
\end{eqnarray}
Since $\mathcal{L}(0)=0$, we observe that the map $\mathcal{L}:\mathcal{B}_R\rightarrow\mathcal{B}_R$
and is a contraction,  having a
unique fixed point -- the trivial one -- if
\begin{eqnarray}
\label{Taylor21}
L(R)<\frac{\delta}{\beta}.
\end{eqnarray}
We consider the polynomial function
\begin{eqnarray}
\label{Taylor22}
\Pi(R):=L(R)-\frac{\delta}{\beta}.
\end{eqnarray}
A threshold value for the existence of nontrivial breather solutions
can be derived from condition (\ref{Taylor21}): the polynomial
equation $\Pi(R)=0$, has exactly two real roots $R^*<0<R_*$, such that
$R_*=-R^*$. Thus
$$
\Pi(R)<0\;\;\mbox{for every}\;\;R\in(0,R_*),
$$
that is, condition (\ref{Taylor21}) is satisfied if $R\in (0, R_*)$.
We summarize our results in the following
\begin{theorem}\label{notrisat}
We assume that the parameters $\epsilon>0$ and $\beta,\Omega>0$ are
given such that
\begin{eqnarray*}
\beta>\Omega.
\end{eqnarray*}
Let $R_*>0$ denote the unique positive root of the polynomial equation
$\Pi(R)=0$, where $\Pi(R)$ and $L(R)$ are given by (\ref{Taylor22})
and (\ref{Taylor17}) respectively.  Then a breather solution (\ref{posfreq})
of (\ref{DNLSsn}), must have power bigger than
\begin{eqnarray}
\label{fplb}
\mathcal{P}_{\beta,\Omega}:=R_*^2(\beta,\Omega)
\end{eqnarray}
\end{theorem}

\paragraph{\textbf{Cubic-Quintic approximation}} The saturable nonlinearity can
be approximated by a cubic-quintic approximation ($m=2$ in Taylor's
formula (\ref{Taylor1})).

\begin{figure}
\begin{center}
    \begin{tabular}{cc}
    \includegraphics[scale=0.33]{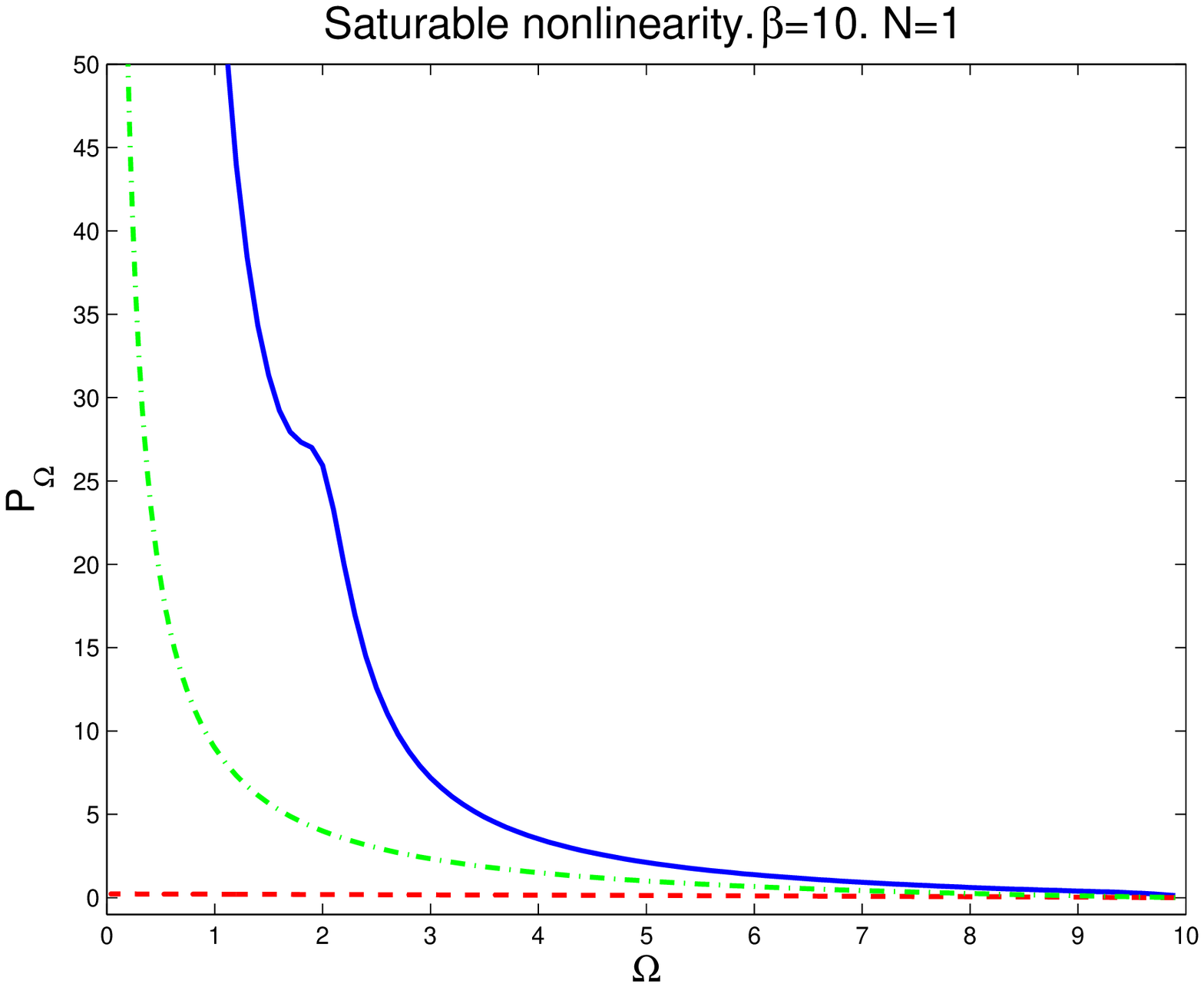} &
    \includegraphics[scale=0.33]{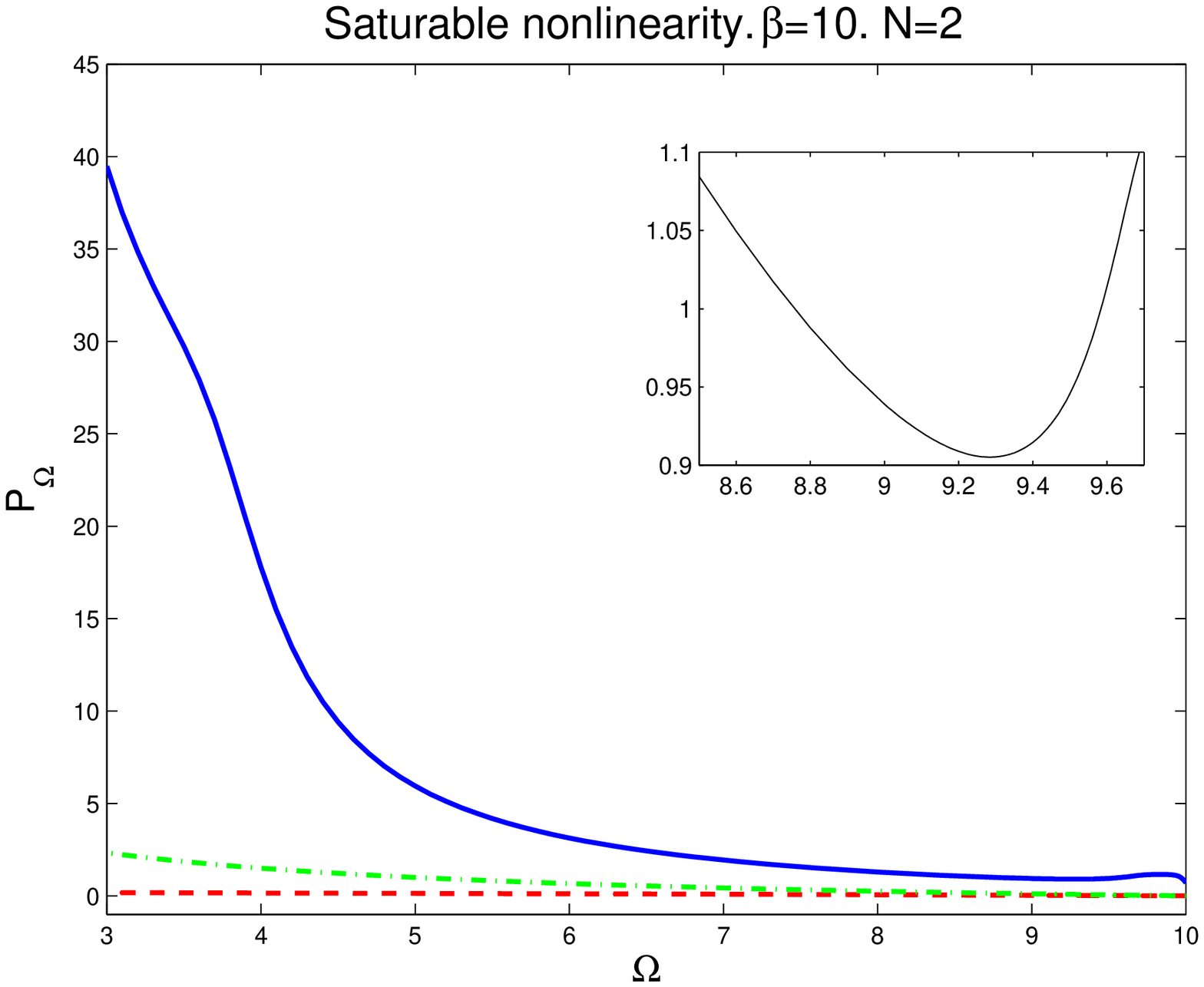}
    \end{tabular}
    \caption{Dependence of the power with respect to the frequency for
      $1D$ and $2D$ lattices of the focusing DNLS with saturable
      nonlinearity for $\beta=10$ and $\epsilon=1$. Power decreases
      approaching the lower bounds when $\Omega$ tends to $\beta$.
      Increased accuracy of the variational lower bound (\ref{HIIt3}),
      for $N=1$. An excitation threshold appears in the 2D-lattice.}
    \label{fig:saturable2}
\end{center}
\end{figure}

We consider first the approximation without taking into account the
remainder term (the term $L_2(R)$ does not appear in the polynomial
equation).  Stationary wave solutions (\ref{posfreq}) satisfy the
infinite system of algebraic equations
\begin{eqnarray}
\label{qubic1}
 -\epsilon(\Delta_d\phi)_n-\Omega\phi_n=-\beta (1-|\phi_n|^2
+|\phi_n|^4)\phi_n,\;\;n\in\mathbb{Z}^N,\;\;\beta>\Omega>0.
 \end{eqnarray}
 In this case, the threshold value is $\mathcal{P}_{\beta,\Omega} =
 R_{*}^2$, where $R_*$ is the root of the quadratic equation
 $\Pi(R)=3R^2+5R^4-\frac{\delta}{\beta}=0$, $\delta=\beta-\Omega$,
\begin{eqnarray}
\label{qubic2}
\mathcal{P}_{\beta,\Omega}={R_{*}}^2(\beta,\Omega) =
-\frac{3}{10}+\frac{(29\beta-20\Omega)^{1/2}}{10\beta^{1/2}}.
\end{eqnarray}
Setting, for example $\beta=2$, $\Omega=0.5$, we obtain the threshold value
$$
\mathcal{P}_{2,0.5}\approx 0.189898.
$$
\paragraph{\textbf{Exact saturable nonlinearity}}
For the exact saturable nonlinearity we should take into account the
remainder term: we look for the root $R^*$, of the equation
$\Pi(R)=3R^2+5R^4+7R^6+2R^8-\frac{\delta}{\beta}=0$. For $\beta=1$,
$\Omega=0.5$, we obtain
$$\mathcal{P}_{2,0.5}\approx 0.180917.$$
Note that the threshold value appears to be the same for parameters
$\beta>\Omega>0$, giving the same ratio $\frac{\beta-\Omega}{\beta}$.
\paragraph{\textbf{Finite dimensional lattice}}
We may also derive a threshold value, taking into account the finite
dimensionality of the lattice, when the problem is supplemented with
Dirichlet boundary conditions. We may replace the constant
$\delta=\beta-\Omega>0$ by the constant
$$
\delta_1=\epsilon\lambda_1+\beta-\Omega,
$$
where $\beta>\Omega>0$.  Therefore, in this case one has to work with
the polynomial equation
\begin{eqnarray}
\label{Taylor22f}
\Pi(R):=L(R)-\frac{\delta_1}{\beta}.
\end{eqnarray}
For the case of the cubic quintic approximation, the threshold value
(\ref{qubic2}), (without considering the remainder term), becomes
\begin{eqnarray}
\label{qubic3}
\mathcal{P}_{\beta,\Omega, D}=-\frac{3}{10}+\frac{[29\beta
+20(\epsilon\lambda_1-\Omega)]^{1/2}}{10\beta^{1/2}},
\end{eqnarray}
involving the principal eigenvalue $\lambda_1>0$ of the operator
$-\Delta_d$ and lattice spacing $\epsilon$.
\begin{figure}
\begin{center}
    \begin{tabular}{cc}
    \includegraphics[scale=0.33]{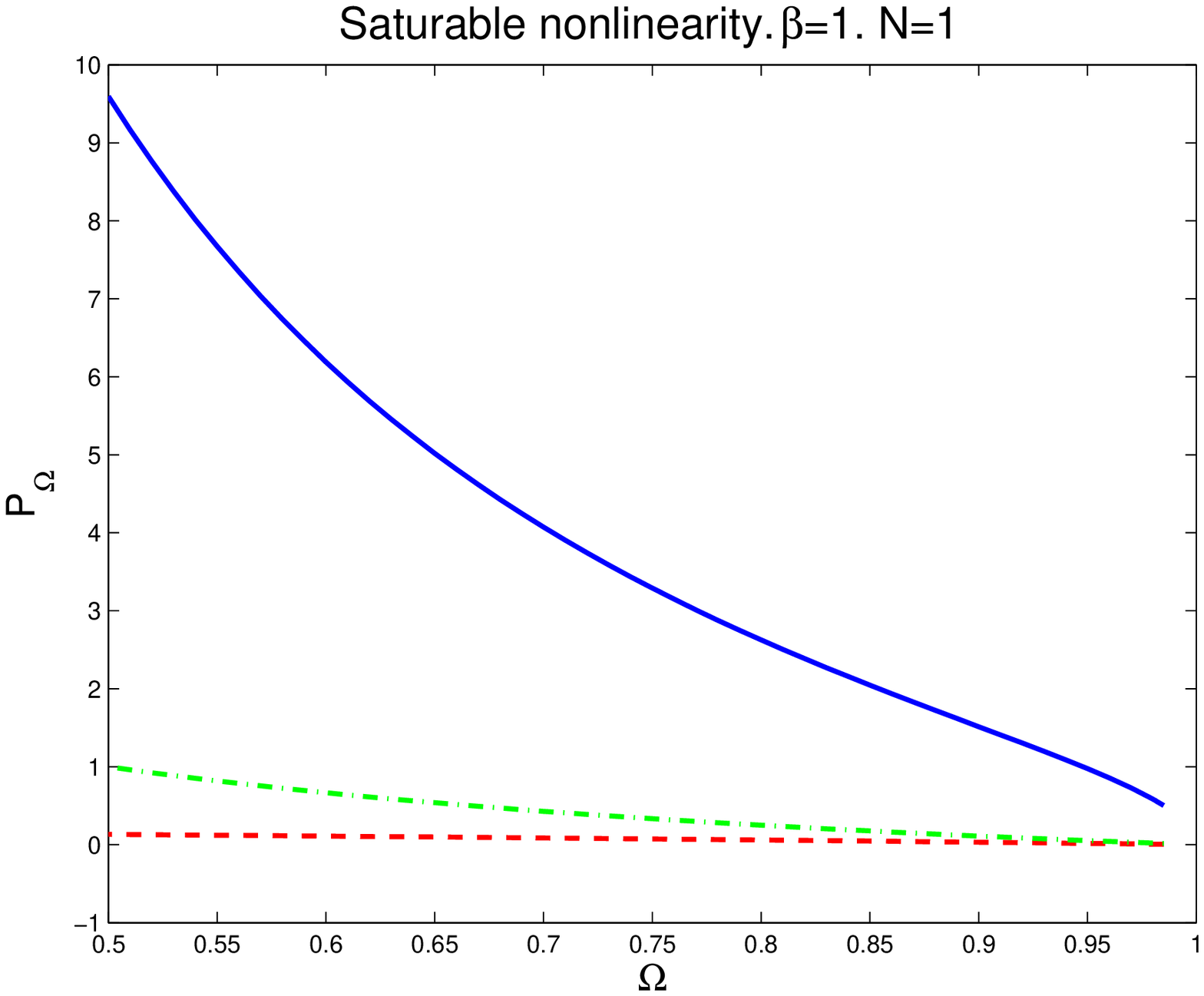} &
    \includegraphics[scale=0.33]{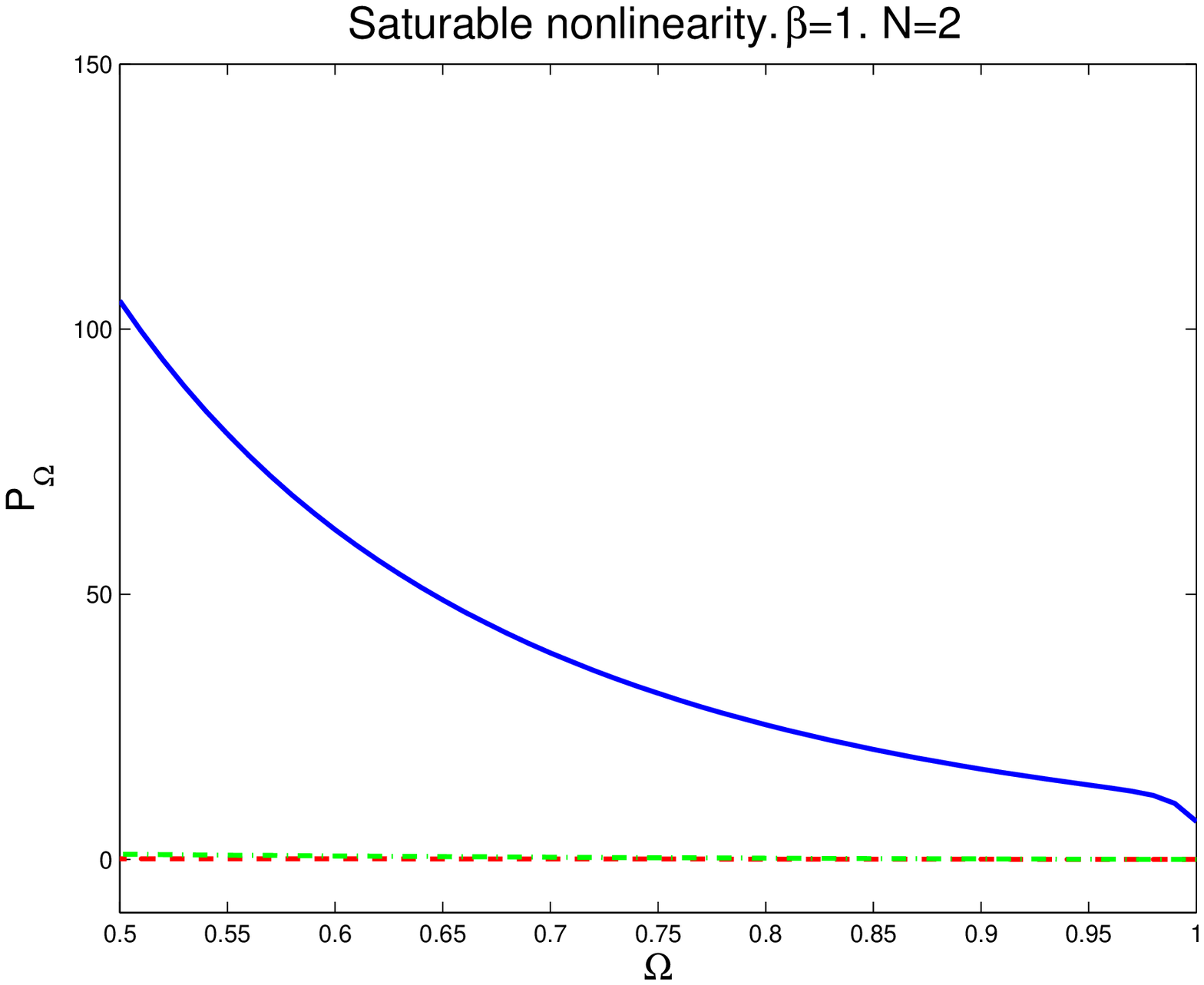}
    \end{tabular}
    \caption{Dependence of the power with respect to the frequency for
      $1D$ and $2D$ lattices of the focusing DNLS with saturable
      nonlinearity for $\beta=1$ and $\epsilon=1$.  Power decreases
      approaching the theoretical lower bounds when $\Omega$ tends to
      $\beta$.}
    \label{fig:saturable1}
\end{center}
\end{figure}
\paragraph{\textbf{Numerical study.}} We performed numerical studies to test
the lower bound (\ref{HIIt3}) derived in Theorem \ref{exomega}, and
the threshold (\ref{fplb}) derived in Theorem \ref{notrisat}.  In
figures \ref{fig:saturable1} and \ref{fig:saturable2}, we show the
dependence of the power $\mathcal{P}_\Omega=\sum|\psi_n|^2$ with
respect to the frequency $\Omega$. In the figures, the solid line
represents the numerically calculated power, while the dashed line
corresponds to the analytical threshold defined as the root of the
equation (\ref{Taylor22f}) for the finite with Dirichlet b.c.
lattice. The dot-dashed line corresponds to the lower bound defined in
(\ref{HIIt3}). In most of the cases considered in our study, the
difference between infinite and finite lattices is difficult to see,
so the former has not been represented in the figures. As we cannot
model an infinite lattice numerically, the numerical results for the
power correspond to finite lattices with Dirichlet b.c. Note that, due
to the resonance with linear modes, the frequencies of the breathers are
limited by the condition $\Omega<\beta$. Besides, the continuation of
the solutions is quite difficult close to this limit.

Figure \ref{fig:saturable1} refers to the parameters $\beta=1$ and
$\epsilon=1$.  The figure verifies that we should not expect existence
of breather solutions below the threshold value (\ref{fplb}). As
$\Omega$ increases to the limit $\beta$, power decreases, approaching
both theoretical estimates. We also note that the lower bound
(\ref{HIIt3}) predicts the decrease of the power as frequency
increases, approaching the threshold (\ref{fplb}).

Figure \ref{fig:saturable2} considers the case $\beta=10$,
$\epsilon=1$ in 1D and 2D-lattices. We observe the increased
accuracy of  the qualitative and quantitative predictions of the
variational lower bound (\ref{HIIt3}), in the 1D-case. In the
2D-case an excitation threshold appears, i.e.\ there exist a 
minimum of the power (excitation threshold), and as frequency
increases, the power increases, reaching a local maximum. This
behaviour is in accordance with that described in the recent work
\cite{RodMagnus}. In the focusing saturable DNLS, this behaviour
seems to appear in the 2D-lattice for larger values of the
parameter $\beta$, while in the defocusing case this behaviour seems
to appear when only the dimension of the lattice is increased.

The ``limiting case'' with respect to the size of $\beta$, $\beta=10$,
clearly demonstrates that (\ref{fplb}) is a quite sharp estimate of
the threshold value on the power, for the existence of a breather in
the focusing saturable nonlinearity. The corresponding breathers with
large frequencies are  real examples demonstrating that this estimate
is the smallest value on the power of an existing breather. The same
closeness appears for the variational lower bound (\ref{HIIt3}).

\section{DNLS equation with power nonlinearity}
\label{DNSss}
\subsection{The defocusing DNLS $\beta>0$:
solutions $\psi_n(t)=e^{-\mathrm{i}\Omega t}\phi_n, \Omega>0$.}
In this subsection we derive a threshold on the power for the existence
of non-trivial breather solutions for the DNLS equation with power
nonlinearity (\ref{DNLSsn}), in the {\em defocusing case} $\beta>0$.  We
seek breather solutions
\begin{eqnarray}
\label{posfreqsigma}
\psi_n(t)=e^{-\mathrm{i}\Omega t}\phi_n,\;\;\Omega>0,
\end{eqnarray}
for the DNLS equation
\begin{eqnarray}
\label{DNLSsL}
\mathrm{i}\dot{\psi}_n+\epsilon(\Delta_d\psi)_n-
\beta|\psi_n|^{2\sigma}\psi_n=0,\;\;\beta>0.
\end{eqnarray}
Solutions (\ref{posfreqsigma}) of DNLS equation (\ref{DNLSsL}),
satisfy the equation
\begin{eqnarray}
\label{EuLag2p}
-\epsilon(\Delta_d\phi)_n-\Omega\phi_n=-\beta|\phi_n|^{2\sigma}\phi_n,
\;\;\Omega>0,\;\beta>0.
\end{eqnarray}
We consider the Hamiltonian for the defocusing DNLS (\ref{DNLSsL})
\begin{eqnarray}
\label{Hamsigma}
\mathcal{H}_{\sigma}[\phi]=\epsilon\sum_{\nu=1}^{N}||\nabla^+_{\nu}
\phi||_{2}^2+\frac{\beta}{\sigma+1}\sum_{n\in\mathbb{Z}^N}
|\phi_n|^{2\sigma+2},\;\;\beta>0,
\end{eqnarray}
and the energy functional
\begin{eqnarray}
\label{enegsigma}
\mathcal{E}_{\Omega}[\phi]=\epsilon\sum_{\nu=1}^{N}||\nabla^+_{\nu}
\phi||_{2}^2-\Omega\sum_{n\in\mathbb{Z}^N}|\phi_n|^{2},\;\;\Omega>0.
\end{eqnarray}
We also recall that the derivative of the functional
\begin{eqnarray*}
\mathcal{L}_R[\phi]=\sum_{|n|\leq K}|\phi_n|^{2\sigma +2},
\end{eqnarray*}
is given by (see also \cite[Lemma 2.3]{K1})
\begin{eqnarray}
\label{derivsigma}
\left<\mathcal{L}_R'[\phi],\psi\right>=(2\sigma+2)\mathrm{Re}
\sum_{|n|\leq K}|\phi_n|^{2\sigma}\phi_n\overline{\psi},\;
\mbox{for all}\;\;\psi\in\ell^2(\mathbb{Z}^N_K).
\end{eqnarray}
Existence of a nontrivial breather solution (\ref{posfreqsigma}), will be
derived by considering constrained minimization problems similar to
those in Section 3 (see also \cite{Wein99}). We have the following
\begin{theorem}
\label{exfrsigma}
$\mathrm{A}$. Consider the variational problem on $\ell^2(\mathbb{Z}^N_K)$
\begin{eqnarray}
\label{infsigma}
\inf\left\{\mathcal{H_\sigma}[\phi]\;:\mathcal{P}[\phi]=R^2>0\right\},
\end{eqnarray}
for some $\beta>0$. There exists a minimizer
$\phi^*\in\ell^2(\mathbb{Z}^N_K)$ for the variational problem
(\ref{infsigma}) and $\omega(R)>0$, such that
$\Omega(R)>\epsilon\lambda_1$, with both satisfying the Euler-Lagrange
equation
\begin{eqnarray*}
\label{EL1sigma}
-\epsilon(\Delta_d\phi)_n&+&\beta|\phi_n|^{2\sigma}\phi_n
=\Omega\phi_n,\;\; |n|\leq K,\\
\phi^*_n&=&0,\;\;|n|>K,
\end{eqnarray*}
with $\mathcal{P}[\phi^*]=R^2$.\vspace{0.1cm}
\newline
$\mathrm{B}$. Consider the variational problem on $\ell^2(\mathbb{Z}^N_K)$
\begin{eqnarray}
\label{infsigmaE}
\inf\left\{\mathcal{E}_{\Omega}[\phi]\;:\sum_{|n|\leq K}|\phi_n|^{2\sigma +2}
=M>0\right\},
\end{eqnarray}
for some $\Omega>0$. Assume that
\begin{eqnarray}
\label{negEsigma}
\Omega>4\epsilon N.
\end{eqnarray}
There exists a minimizer $\hat{\phi}\in\ell^2(\mathbb{Z}^N_K)$ for the
variational problem (\ref{infsigmaE}) and $\beta(M)>0$, both
satisfying the Euler-Lagrange equation (\ref{EL1sigma}), and
$\sum_{|n|\leq K}|\hat{\phi}_n|^{2\sigma+2}=M$. Assume that the
power of the minimizer is $\mathcal{P}[\hat{\phi}]=R^2$. Then the
power satisfies the lower bound
\begin{eqnarray}
\label{thresE0}
\left[\frac{\Omega-4N\epsilon}{\beta (\sigma +1)}
\right]^{\frac{1}{\sigma}}\leq R^2:=\mathcal{P}_{1}[\hat{\phi}].
\end{eqnarray}
\end{theorem} {\bf Proof:} $\mathrm{A}$. Relation (\ref{derivsigma})
implies that the functional $\mathcal{H}_{\sigma}:
\ell^2(\mathbb{Z}^N_K) \rightarrow\mathbb{R}$ is a $C^1$-functional.
It is bounded from below, since from (\ref{eigchar})
\begin{eqnarray*}
\mathcal{H}_{\sigma}[\phi]\geq \epsilon\sum_{\nu=1}^{N}
||\nabla^+_{\nu}\phi||_{2}^2\geq \epsilon\lambda_1 R^2.
\end{eqnarray*}
The same variational arguments of Section 3 imply the existence of a
minimizer $\phi^*\in\ell^2(\mathbb{Z}^N_K)$ of $\mathcal{H}_{\sigma}$,
and the Lagrange multiplier $\Omega(R)>0$, such that
\begin{eqnarray}
\label{Hsigma}
\left<\mathcal{H}_{\sigma}'[\phi^*]-
\Omega\mathcal{N}_R'[\phi],\psi\right>&=&
2\epsilon\sum_{\nu=1}^N(\nabla_\nu^+\phi^*,\nabla_\nu^+\psi)_{2}
+2\beta\mathrm{Re}\sum_{|n|\leq K}|\phi^*_n|^{2\sigma}
\phi^*_n\overline{\psi}_n\nonumber\\
&&-2\Omega\sum_{|n|\leq K}\phi^*_n\overline{\psi_n}=0,\;\;
\mbox{for all}\;\;\psi\in\ell^2(\mathbb{Z}^N_K).
\end{eqnarray}
Then setting $\psi=\phi^*$ in (\ref{Hsigma}), and by using
(\ref{crucequiv}), we obtain that
\begin{eqnarray}
\label{toth2}
2\epsilon\lambda_1\sum_{|n|\leq K}|\phi^*_n|^{2}\leq
2\epsilon\sum_{\nu=1}^{N}||\nabla^+_{\nu}\phi^*||_{2}^2
+2\beta \sum_{|n|\leq K}|\phi^*_n|^{2\sigma+2}=
   2\Omega\sum_{|n|\leq K}|\phi^*_n|^{2},
\end{eqnarray}
which shows that
$\Omega(R)>\epsilon\lambda_1>0$.

$\mathrm{B}$. The functional $\mathcal{E}_{\Omega}$ is bounded from
below: the equivalence of norms (\ref{fnorms}), implies the existence
of a $N$-dependent constant $C_2$, such that
\begin{eqnarray}
\label{equivsigma}
||\phi||_{2}^2\leq C_2^2||\phi||_{2\sigma+2}^2,\;\;
\mbox{for all}\;\;\phi\in\ell^2(\mathbb{Z}^N_K).
\end{eqnarray}
Then using (\ref{equivsigma}), we find that
\begin{eqnarray}
\mathcal{E}_{\Omega}[\phi]&\geq& -
\Omega\sum_{|n|\leq K}|\phi_n|^2\\
&\geq&-\Omega C_2^2\left(\sum_{|n|\leq K}
|\phi_n|^{2\sigma +2}\right)^{\frac{1}{\sigma+1}}\\
&\geq& -\Omega C_2^2M^{\frac{1}{\sigma+1}}.
\end{eqnarray}
Again the existence of the minimizer $\hat{\phi}$ and of the Lagrange
multiplier $\lambda(M)\in\mathbb{R}$ can be obtained by the same
arguments as in Section 3. Moreover by using (\ref{derivsigma}), we
have that
\begin{eqnarray}
\label{Esigma}
\left<\mathcal{E}_{\Omega}'[\hat{\phi}]-\lambda\mathcal{L}_{R}'
[\hat{\phi}],\psi\right>&=&
2\epsilon\sum_{\nu=1}^N(\nabla_\nu^+\hat{\phi},\nabla_\nu^+\psi)_{2}
-2\Omega\mathrm{Re}\sum_{|n|\leq K}\hat{ \phi}_n\overline{\psi}_n\nonumber\\
&&-2(\sigma +1)\lambda\sum_{|n|\leq K}|\hat{\phi}_n|^{2\sigma}
\hat{\phi}_n\overline{\psi_n}=0,
\end{eqnarray}
for all
$\psi\in\ell^2(\mathbb{Z}^N_K)$. Setting $\psi=\hat{\phi}$ in (\ref{Esigma}), we obtain
\begin{eqnarray}
\label{Esigma2}
2\mathcal{E}_{\Omega}[\hat{\phi}]=2\epsilon\sum_{\nu=1}^{N}
||\nabla^+_{\nu}\hat{\phi}||_{2}^2-2\Omega\sum_{|n|\leq K}|
\hat{\phi}_n|^2=2(\sigma +1)\lambda\sum_{|n|\leq K}|\hat{\phi}|^{2\sigma+2}.
\end{eqnarray}
We observe that
\begin{eqnarray}
\label{Esigma3}
\mathcal{E}_{\Omega}[\hat{\phi}]\leq 4\epsilon N
\sum_{|n|\leq K}|\hat{\phi}_n|^2-\Omega\sum_{|n|\leq K}|\hat{\phi}_n|^2.
\end{eqnarray}
Thus $\mathcal{E}_{\Omega}[\hat{\phi}]<0$ if (\ref{negEsigma}) is
satisfied. Note that due to the estimate (\ref{prineig}), the
condition (\ref{negEsigma}) implies  that
\begin{eqnarray}
\label{negEsigma1}
\Omega>\epsilon\lambda_1.
\end{eqnarray}
Then assuming (\ref{negEsigma}), we find that $\lambda(M)<0$. We set
$\lambda=-\beta,\,\beta>0$.  Finally, we assume that the power of the
nontrivial minimizer $\hat{\phi}$ is $\sum_{|n|\leq
  K}|\hat{\phi}_n|^2=R^2$. Then, returning to (\ref{Esigma2}), and by
using (\ref{lp1}) which holds also in the finite dimensional lattice,
we get
\begin{eqnarray}
\label{thresE}
2\Omega R^2&=&2\epsilon\sum_{\nu=1}^{N}||\nabla^+_{\nu}
\hat{\phi}||_{2}^2+2(\sigma +1)\beta\sum_{|n|\leq K}|
\hat{\phi}_n|^{2\sigma+2}\nonumber\\
&&\leq 8\epsilon N\sum_{|n|\leq K}|\hat{\phi}_n|^2
+2(\sigma+1)\beta\sum_{|n|\leq K}|\hat{\phi}_n|^{2\sigma+2}\nonumber\\
&&\leq8\epsilon N\sum_{|n|\leq K}|\hat{\phi}_n|^2
+2(\sigma +1)\beta\left(\sum_{|n|\leq K}|
\hat{\phi}_n|^2\right)^{\sigma+1}\nonumber\\
&&\leq 8\epsilon N R^2+2(\sigma +1)\beta R^{2\sigma+2}.
\end{eqnarray}
Under condition (\ref{negEsigma}), inequality (\ref{thresE}) implies
the lower bound (\ref{thresE0}). \ 
$\diamond$

A threshold for the power could be also derived from the case
$\mathrm{A.}$ of Theorem \ref{exfrsigma}: working exactly as for the
derivation of that in (\ref{thresE}), we find from (\ref{toth2}), that
\begin{eqnarray}
\label{thresE2a}
\Omega R^2\leq 4\epsilon N R^2+\beta R^{2\sigma +2}.
\end{eqnarray}
Thus, in the case of Theorem \ref{exfrsigma} $\mathrm{A}$, and under the
hypothesis that the Lagrange multiplier $\Omega(R)$ of the case
$\mathrm{A}$ is taking values $\Omega>4\epsilon N$, we find from
(\ref{thresE2a}) that
\begin{eqnarray}
\label{thresE2b}
\left[\frac{\Omega-4N\epsilon}{\beta}\right]^{\frac{1}{\sigma}}
\leq R^2:=\mathcal{P}_{2}[\phi^*].
\end{eqnarray}
Comparing with (\ref{thresE0}), it readily follows that
\begin{eqnarray}
\label{relp1p2}
\mathcal{P}_1[\hat{\phi}]<\mathcal{P}_2[\phi^*].
\end{eqnarray}
\subsection{Numerical study for the defocusing DNLS with power
  nonlinearity.}  Similarly to the results of Section 3, for the
saturable DNLS, it seems interesting to test the behaviour of the lower
bounds (\ref{thresE0}) and (\ref{thresE2b}), as thresholds for the
existence of breather solutions for the defocusing DNLS
(\ref{DNLSsL}).  Theorem \ref{exfrsigma} $\mathrm{A}$ implies that,
for a given $\beta>0$, the existence of a frequency
$\Omega>\epsilon\lambda_1$ (as a Lagrange multiplier), and of a
nontrivial minimizer $\phi^*$ of the Hamiltonian (\ref{Hamsigma}),
such that the corresponding breather solution
$\psi_n(t)=e^{-\mathrm{i}\Omega t}\phi_n^*$ has a power satisfying the
lower bound (\ref{thresE2b}), in the case where $\Omega$ is assumed to
be such that $\Omega>4\epsilon N$.  On the other hand, Theorem
\ref{exfrsigma} $\mathrm{B}$ implies, for a given $\Omega>4\epsilon N$,
the existence of $\beta>0$ and of a nontrivial minimizer $\hat{\phi}$
of the energy functional (\ref{enegsigma}), such that the
corresponding breather solution $\psi_n(t)=e^{-\mathrm{i}\Omega
  t}\hat{\phi}_n$, has a power satisfying the lower bound
(\ref{thresE0}).

\begin{figure}
\begin{center}
    \begin{tabular}{cc}
    \includegraphics[scale=0.33]{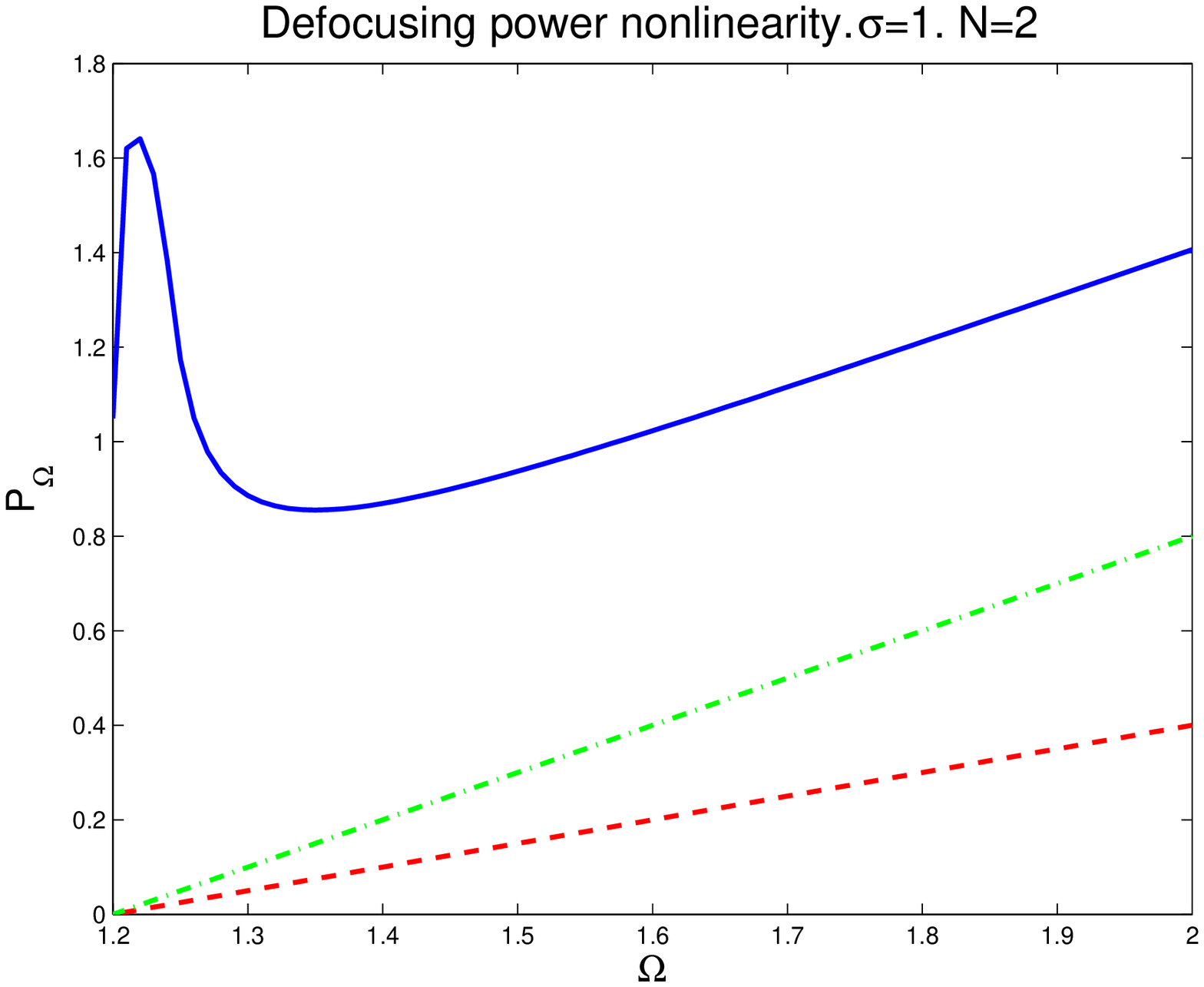} &
    \includegraphics[scale=0.33]{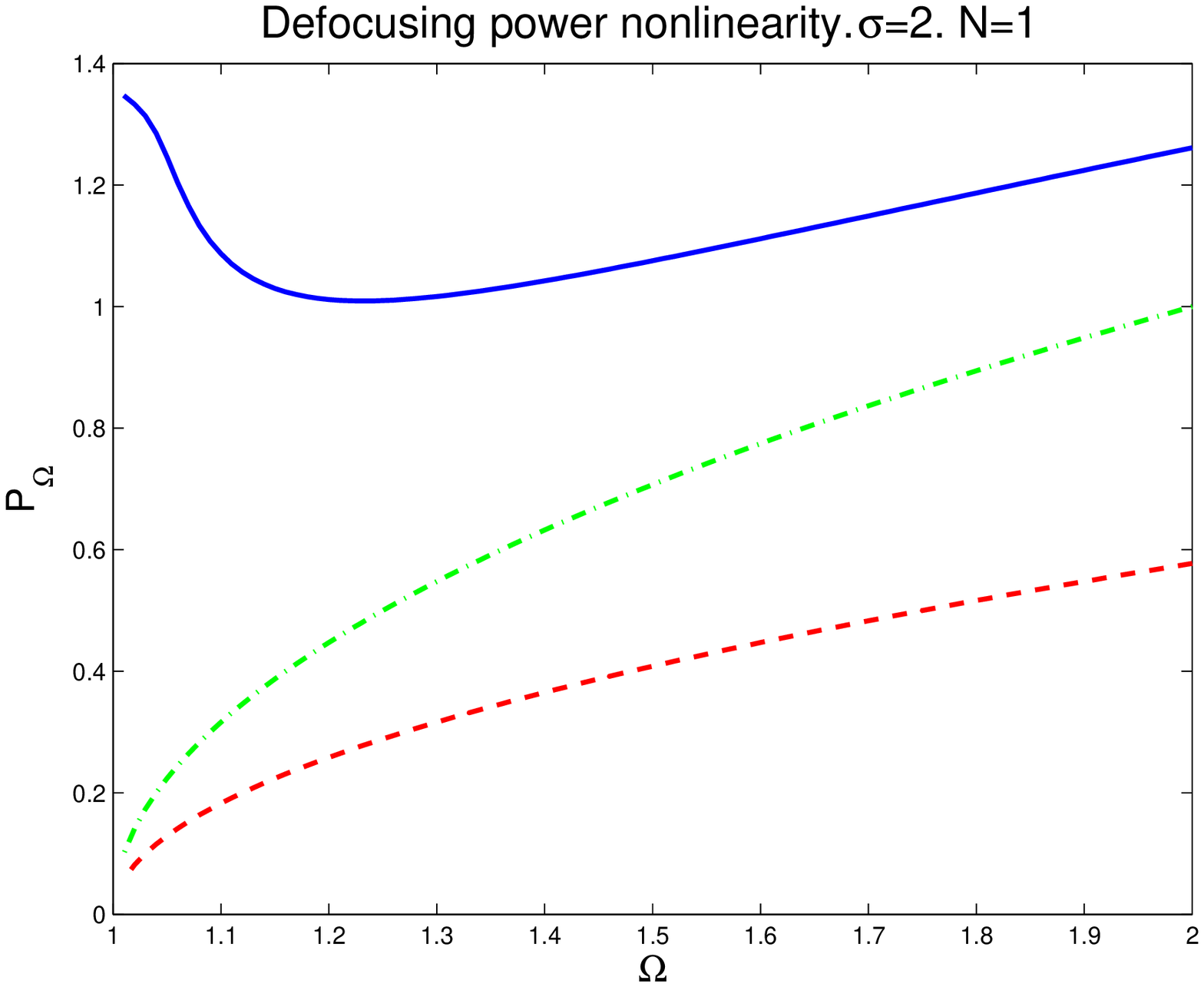}
    \end{tabular}
    \caption{Numerical power for the defocusing DNLS with power
      nonlinearity and (a) $\sigma=1$, $N=2$, $\epsilon=0.15$ (b)
      $\sigma=2$, $N=1$, $\epsilon=0.25$.  The lower bounds
      (\ref{thresE0}) and (\ref{thresE2b}), dot-dashed and dashed lines
      respectively, serve as thresholds for the existence of
      nontrivial breather solutions.}
    \label{fig:defocsigma12}
\end{center}
\end{figure}

\begin{figure}
\begin{center}
    \includegraphics[scale=0.33]{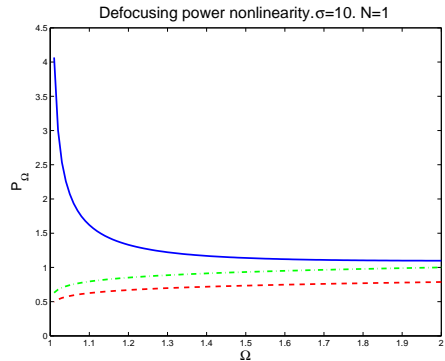}
    \caption{Numerical power for the defocusing DNLS with power
      nonlinearity and $\sigma=10$, $N=1$, $\epsilon=0.25$.  Both
      lower bounds (\ref{thresE0}) and (\ref{thresE2b}), dot-dashed and dashed
      lines respectively, are fulfilled as threshold values.}
    \label{fig:defocsigma10}
\end{center}
\end{figure}

\begin{figure}
\begin{center}
    \begin{tabular}{cc}
    \includegraphics[scale=0.33]{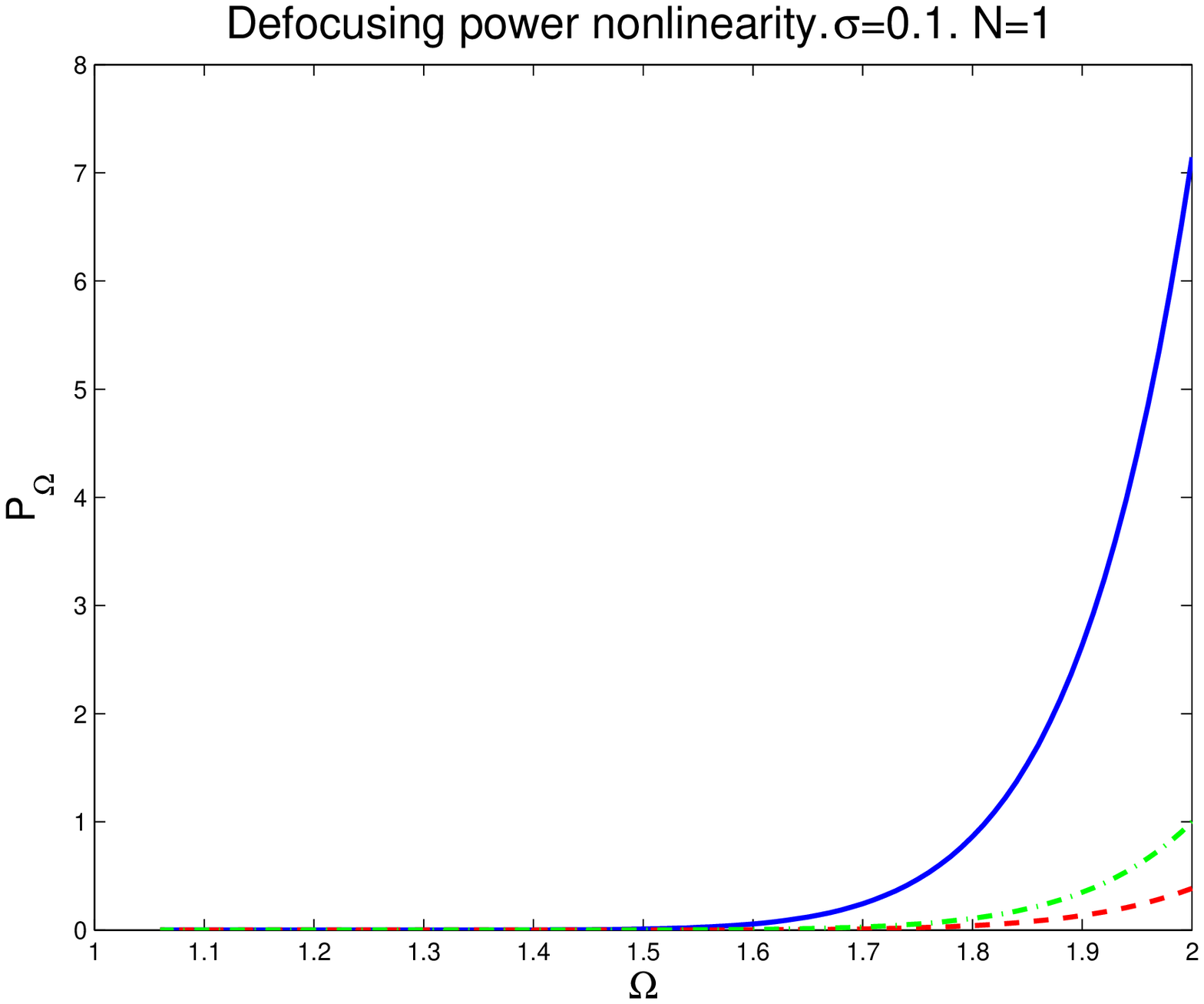} &
    \includegraphics[scale=0.33]{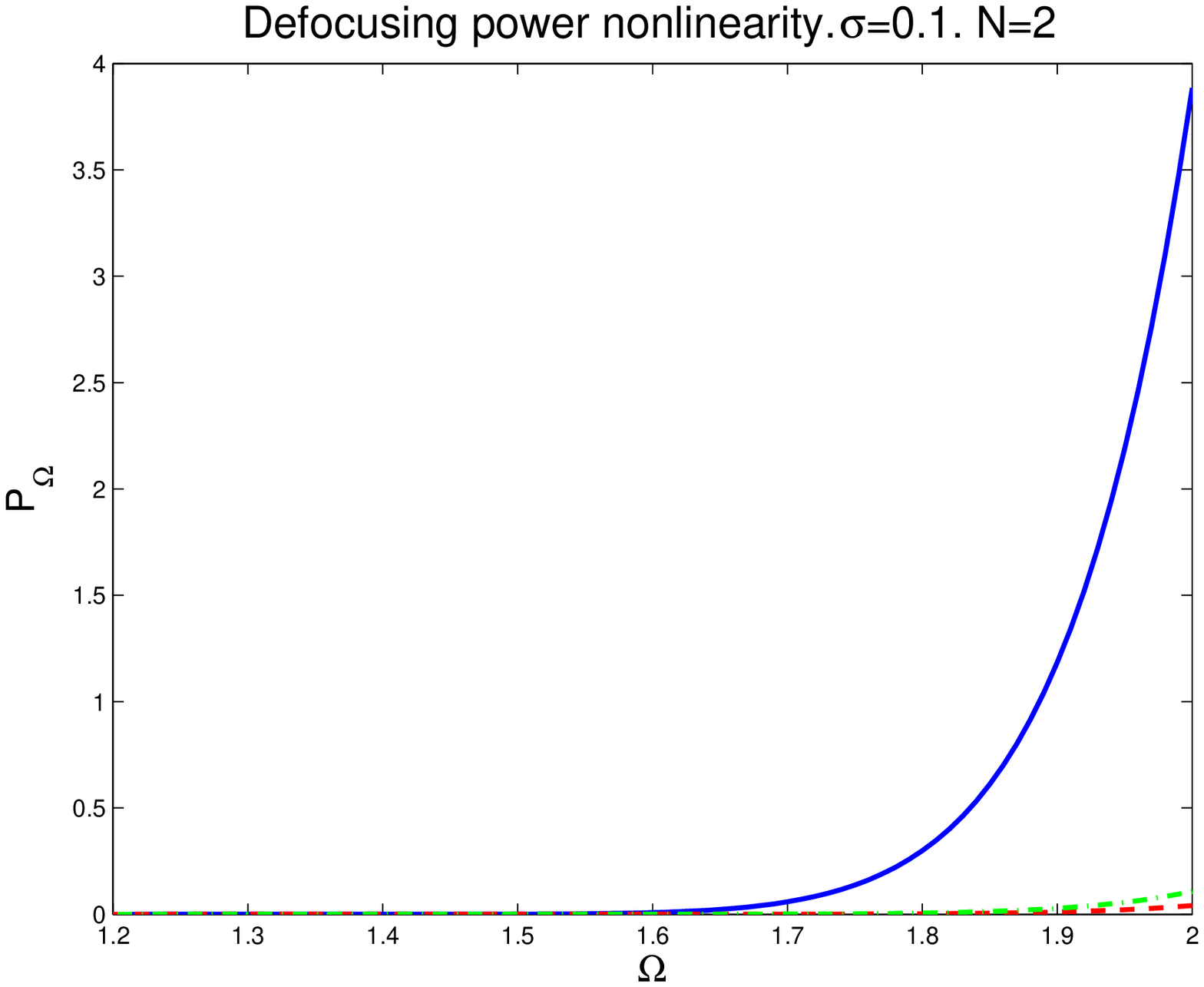}
    \end{tabular}
    \caption{Numerical power for the defocusing DNLS with power nonlinearity case for (a)
      $\sigma=0.1$, $N=1$, $\epsilon=0.25$ (b) $\sigma=0.1$, $N=2$, $\epsilon=0.15$. The lower bounds (\ref{thresE0})
      and (\ref{thresE2b}), dot-dashed and dashed lines respectively,  are
      fulfilled as threshold values.}
    \label{fig:defocsigma01}
\end{center}
\end{figure}

\begin{figure}
\begin{center}
    \includegraphics[scale=0.33]{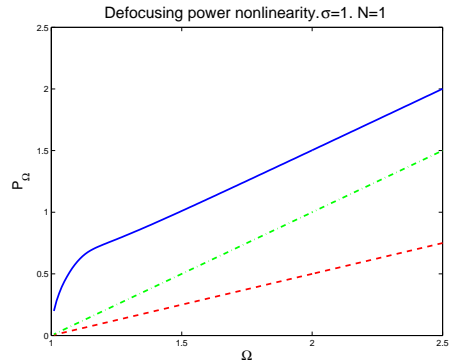}
    \caption{Numerical power for the defocusing DNLS with power nonlinearity case $\sigma=1$, $N=1$, $\epsilon=0.25$.
 The lower bounds (\ref{thresE0})
      and (\ref{thresE2b}), dot-dashed and dashed lines respectively,  are
      fulfilled as threshold values.}
    \label{fig:defocsigma1N1}
\end{center}
\end{figure}

The numerical study is for 1D and 2D-lattices. For the case $N=1$ we
consider $\epsilon=0.25$, and for the case $N=2$ we consider
$\epsilon=0.15$. We study first values of $\sigma$ satisfying
$\sigma\geq\frac{2}{N}$.  In Figure \ref{fig:defocsigma12}, the study
refers to the cases $\sigma=1,\,N=2$, and $\sigma=2,\,N=1$
respectively.  We observe first that the numerical power of the
solutions fulfils both lower bounds (\ref{thresE2b}), (\ref{thresE0}).
Moreover we observe that they can be considered also as {\em
  thresholds} on the power of periodic solutions with frequencies
$\Omega>4\epsilon N$. This fact is revealed by the case $\sigma=10$,
$N=1$ for which the lower bound (\ref{thresE0}) is proved to be a
quite sharp estimate of the power for large frequencies.
The same satisfactory accuracy of the theoretical estimates
(\ref{thresE0}) and (\ref{thresE2b}) is observed also in case of
$\sigma=10$, $N=1$ considered in Figure \ref{fig:defocsigma10}.
We note that the phonon band of the defocusing DNLS equation extends
to the interval $[0,4\epsilon N]$. Then breathers frequencies must lie
in the intervals $\Omega>4\epsilon N$, or $\Omega<0$.  It is the
former case which we consider in this paragraph.  The numerical
studies for the case $\sigma<2/N$ where the excitation threshold
\cite{Wein99} do no exist, seem to fully justify the argument that the
lower bounds (\ref{thresE2b}), (\ref{thresE0}) can be used as
thresholds on the power for the existence of breather solutions. The
results for $\sigma<2/N$ are demonstrated in Figures
\ref{fig:defocsigma01} and \ref{fig:defocsigma1N1}. Figure
\ref{fig:defocsigma01} considers the case $\sigma=0.1$ for $N=1$ and
$N=2$ and (\ref{thresE2b}), (\ref{thresE0}) are clearly sharp
estimates of the smallest value of the power a breather solution can
have. This is justified by the numerical power of breather solutions
with small frequencies. The numerical power for the case $\sigma=1,
N=1$ also fulfils the theoretical estimates.

We remark that the approach of minimizing the linear energy
appeared to be useful also in the defocusing DNLS with power
nonlinearity since the lower bound (\ref{thresE0}) provides in
general better quantitative predictions of the numerical power if
compared with (\ref{thresE2b}).

\subsection{Focusing DNLS with power nonlinearity ($\beta<0$):
solutions $\phi_n(t)=e^{-i\Omega t}\phi_n$, $\Omega<0$.}
We conclude with a numerical study of the focusing DNLS with power
nonlinearity. Thus we shall consider solutions with frequencies $\Omega<0$ (the latter case of the phonon band condition). As in Section 3, we use for convenience, the notation
$\beta=-\Lambda,\;\Lambda>0$, and $\Omega=-\omega,\;\omega>0$.

Instead of the threshold estimate \cite[(2.38), pg. 126]{K1} derived by the fixed point argument, for breather solutions
\begin{eqnarray}
\label{posfreqa}
\psi_n(t)=e^{\mathrm{i}\omega t}\phi_n,\;\;\omega>0,
\end{eqnarray}
of the DNLS (\ref{DNLSs}) in {\em infinite lattices},
\begin{eqnarray}
\label{oldo}
E_{\mathrm{min}}^2:=\frac{1}{4}\left[\frac{\omega}
{\Lambda(2\sigma +1)}\right]^{\frac{1}{\sigma}},
\end{eqnarray}
we may derive a possibly improved one: by using (\ref{Taylor13})
(which holds also for an infinite lattice) instead of \cite[(2.44), pg.
127]{K1}, we find that
\begin{eqnarray}
\label{newo}
\mathcal{P}_{\omega}:=\left[\frac{\omega}{\Lambda(2\sigma +1)}
\right]^{\frac{1}{\sigma}}=4E_{\mathrm{min}}^2.
\end{eqnarray}
\vspace{0.2cm} \newline Although (\ref{newo}), holds also for the case
of the Dirichlet boundary conditions, we may derive a threshold value
taking into account the finite dimensionality of the problem: for
solutions (\ref{posfreqa}), we consider the operator
$\mathcal{T}_{\Omega}:\ell^2(\mathbb{Z}^N_K) \rightarrow
\ell^2(\mathbb{Z}^N_K)$, defined as
\begin{eqnarray}
\label{strongop2}
(\mathcal{T}_{\omega}\phi)_{|n|\leq K}&=&
-\epsilon(\Delta_d\phi)_{|n|\leq K}+\omega\phi_n.
\end{eqnarray}
Clearly
\begin{eqnarray}
\label{newcon3}
\delta_1:=\epsilon\lambda_1+\omega>0,
\end{eqnarray}
and by using (\ref{eigchar}), we get that
\begin{eqnarray}
\label{check2}
(\mathcal{T}_{\omega}\phi,\phi)_{2}=\epsilon\sum_{\nu=1}^N
|{\nabla}^+_{\nu}\phi||^2_{2}+\omega ||\phi||^2_2\geq
\delta_1||\phi||^2_{2}\;\;\mbox{for all}\;\;
\phi\in\ell^2(\mathbb{Z}^N_K).
\end{eqnarray}
Then, by using (\ref{Taylor14}), we  may derive the threshold value
\begin{eqnarray}
\label{newof}
\mathcal{P}_{\omega,D}:=\left[\frac{\epsilon\lambda_1
+\omega}{\Lambda(2\sigma +1)}\right]^{\frac{1}{\sigma}}.
\end{eqnarray}

Throughout this study we have fixed $\epsilon=1$.
A first result is that the power of the solutions fulfils
(\ref{newo}), that is, it is higher than
\begin{eqnarray*}
\mathcal{P}_{\omega}:=\left[\frac{\omega}{\Lambda(2\sigma
+1)}\right]^{\frac{1}{\sigma}}
\end{eqnarray*}
We mention first, that regarding the numerical study of the
threshold (\ref{newof}), as in the case of the saturable
nonlinearity,  we have not observed remarkable improvement or
differences in comparison with (\ref{newo}).  Although the threshold
(\ref{newo}) is independent of the dimension, it is interesting to
compare this with the results of \cite{Wein99}, related to the
conditions on existence of excitation threshold which depends on the
dimension and the nonlinearity exponent $\sigma$.
\begin{figure}
\begin{center}
    \begin{tabular}{cc}
    \includegraphics[scale=0.33]{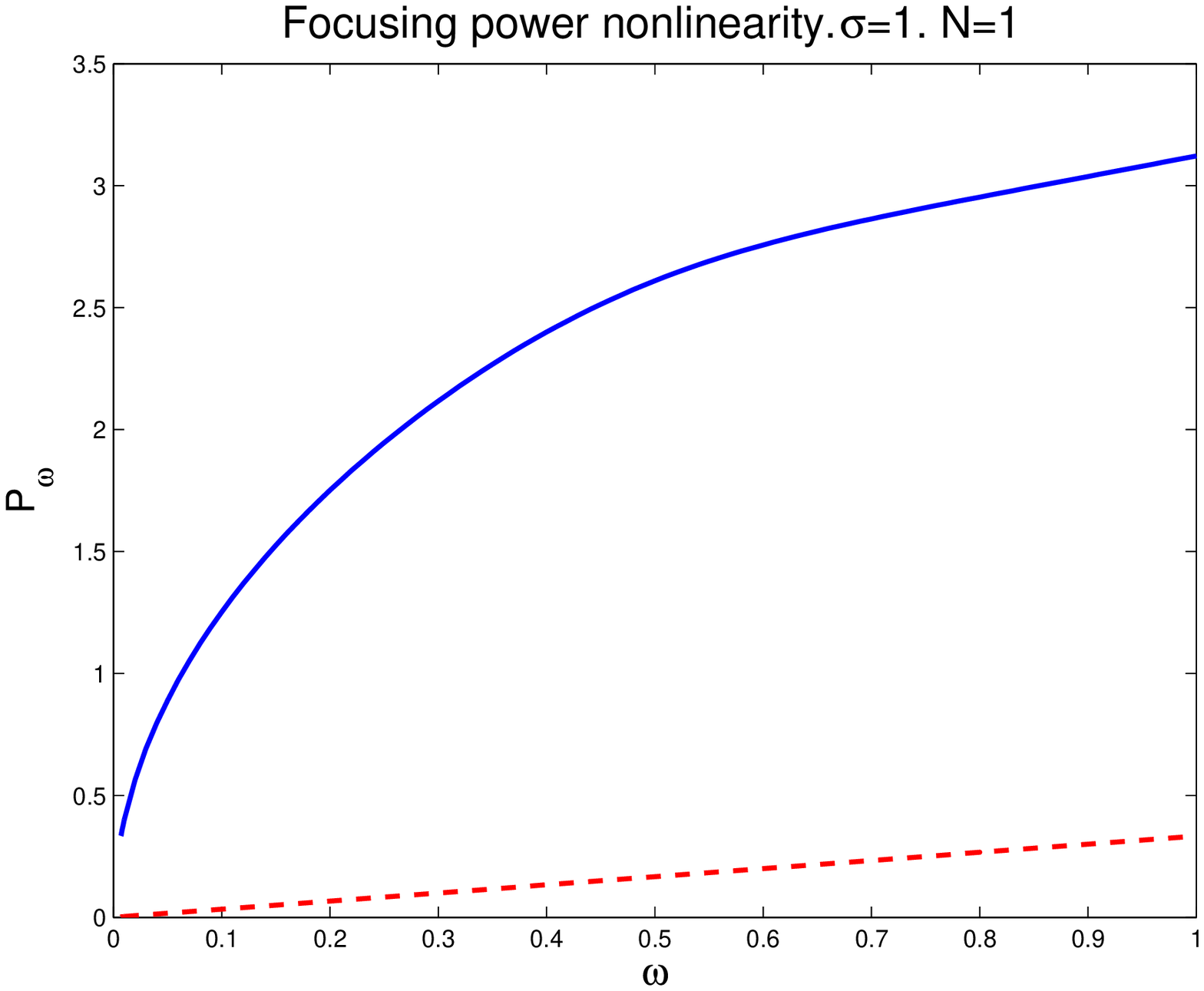} &
    \includegraphics[scale=0.33]{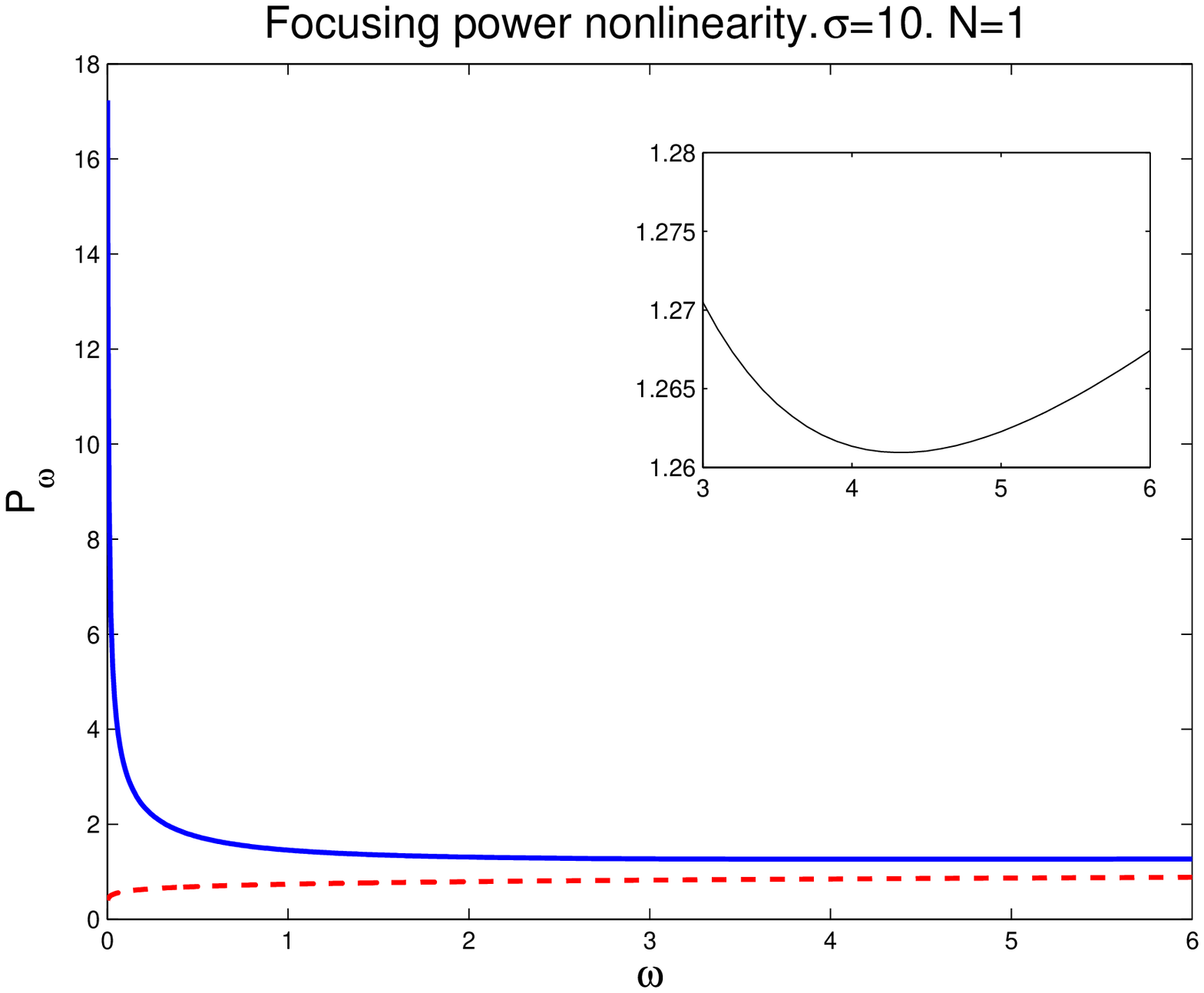}
    \end{tabular}
    \caption{Numerical power for the focusing DNLS with power
      nonlinearity and (a) $\sigma=1$, $N=1$, $\epsilon=1$
      ($\sigma<\frac{2}{N}$-no excitation threshold ), (b)
      $\sigma=10$, $N=1$, $\epsilon=1$
      ($\sigma>\frac{2}{N}$-excitation threshold). The inset in (b)
      shows a magnification of the region where the power reaches its
      minimum value. Dashed line corresponds to lower bound
      (\ref{newof})}
    \label{fig:sigma1}
\end{center}
\end{figure}
\begin{figure}
\begin{center}
    \begin{tabular}{cc}
    \includegraphics[scale=0.33]{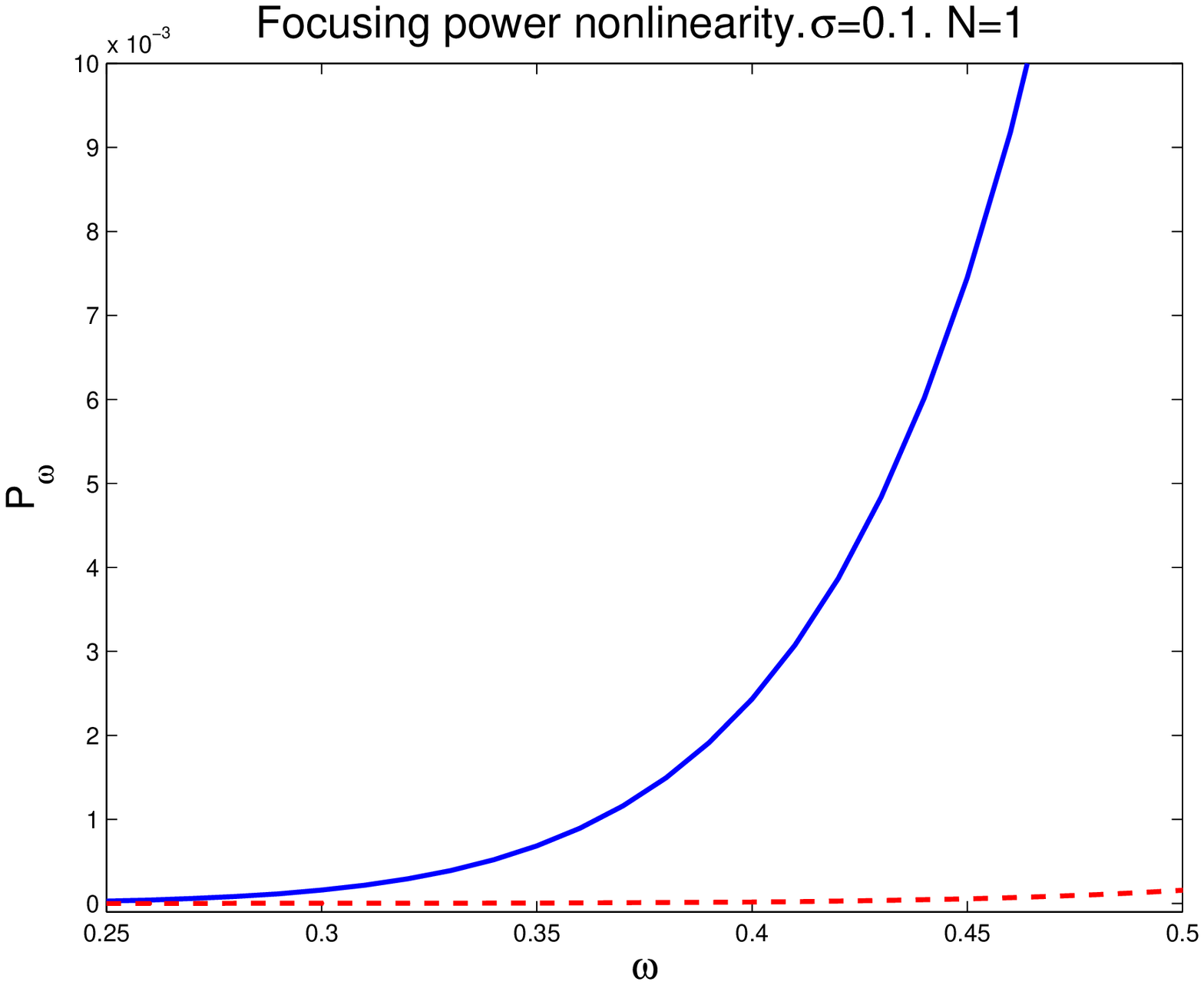} &
    \includegraphics[scale=0.33]{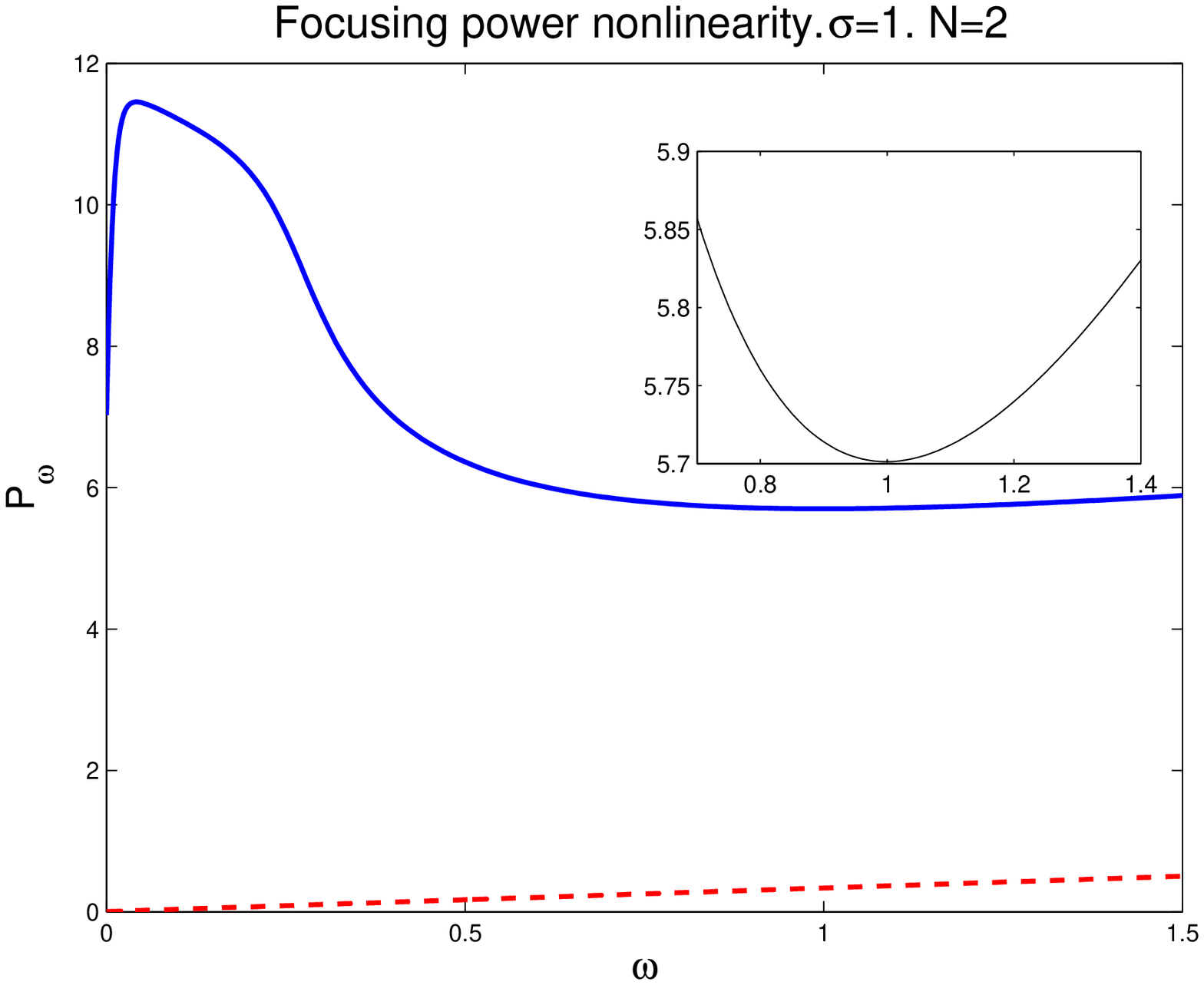}
    \end{tabular}
    \caption{Numerical power for the focusing DNLS with power
      nonlinearity and (a) $\sigma=0.1$, $N=1$, $\epsilon=1$
      ($\sigma<\frac{2}{N}$-no excitation threshold ), (b) $\sigma=1$,
      $N=2$, $\epsilon=1$ ($\sigma=\frac{2}{N}$-excitation threshold).
      The inset in (b) shows a magnification of the region where the
      power reaches its minimum value. Dashed line corresponds to lower
      bound (\ref{newof})}
    \label{fig:sigma10}
\end{center}
\end{figure}
\begin{figure}
\begin{center}
    \begin{tabular}{cc}
    \includegraphics[scale=0.33]{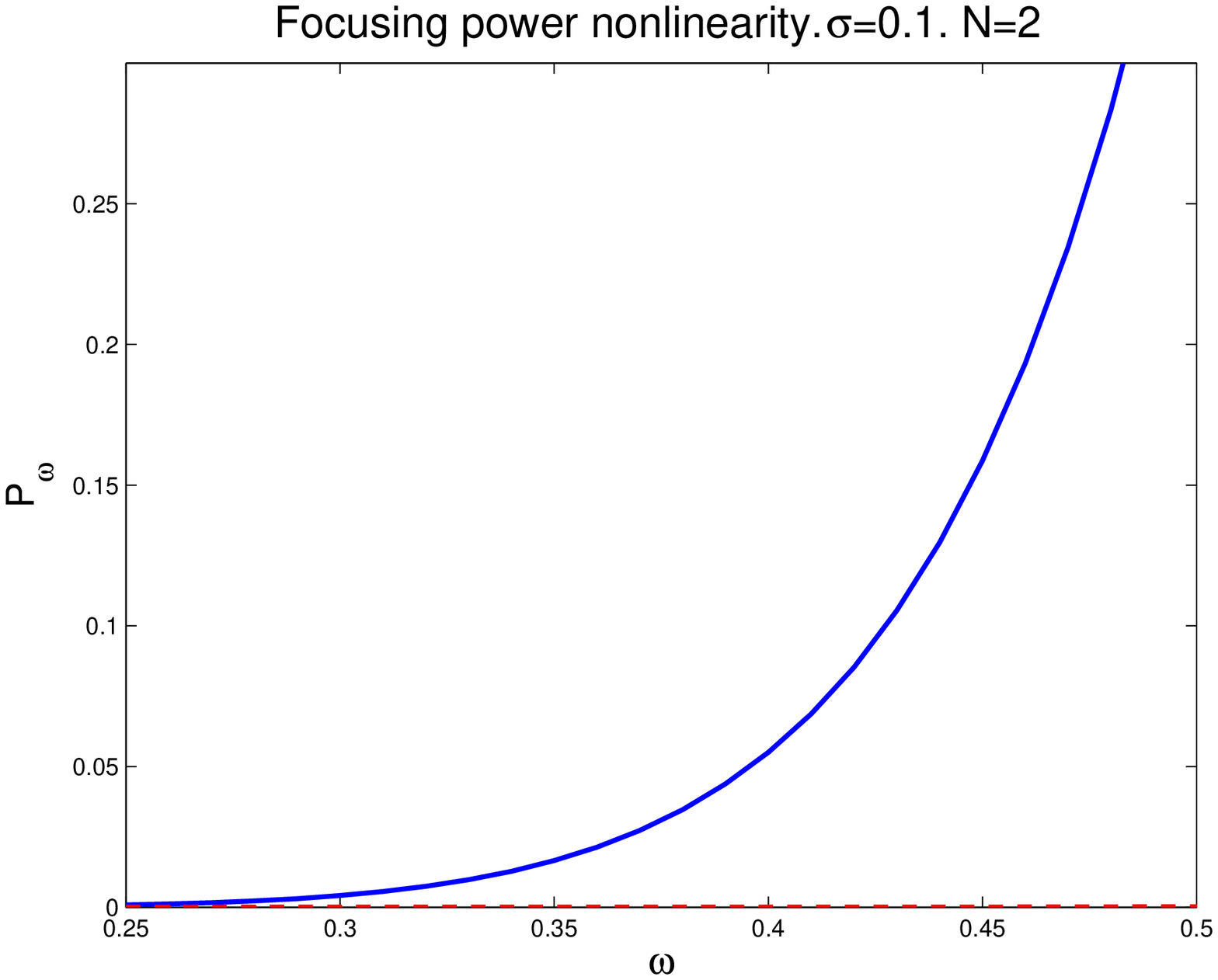} &
    \includegraphics[scale=0.33]{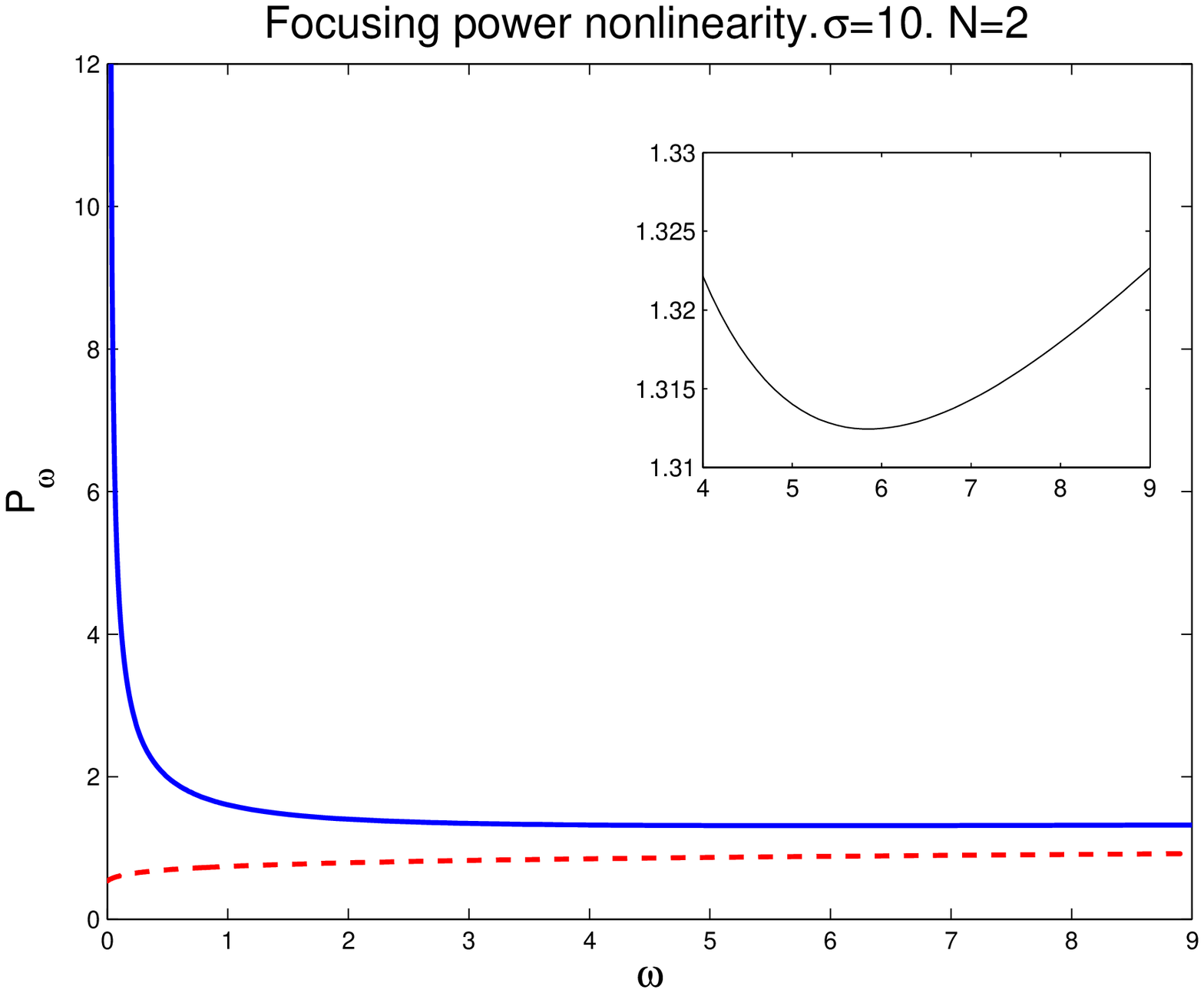}
    \end{tabular}
    \caption{Numerical power for the focusing DNLS with power
      nonlinearity and (a) $\sigma=0.1$, $N=2$, $\epsilon=1$
      ($\sigma<\frac{2}{N}$-no excitation threshold ), (b)
      $\sigma=10$, $N=2$, $\epsilon=1$
      ($\sigma>\frac{2}{N}$-excitation threshold). The inset in (b)
      shows a magnification of the region where the power reaches its
      minimum value. Dashed line corresponds to lower bound
      (\ref{newof})}
    \label{fig:sigma01}
\end{center}
\end{figure}
In Figure \ref{fig:sigma1}, we present the numerical power (solid
curve) for the cases $\sigma=1$ and $\sigma=10$, and $N=1$. The values
$\sigma=1$ and $N=1$ satisfy $\sigma<2/N$, and the power fulfils the
threshold (\ref{newo}) for existence. The case $\sigma=10$ and $N=1$
satisfies $\sigma>2/N$, and as predicted in \cite{Wein99} and
\cite{FlachMac}, the power approaches a minimum (the excitation
threshold). The numerical power still fulfils the threshold
(\ref{newo}). In the latter case, the corresponding breathers with
large frequencies provide real examples demonstrating that
(\ref{newo}) is a quite sharp estimate of the smallest value of the
power a breather can have when $\sigma\geq 2/N$.  Note that the
dependence of this excitation threshold with respect to $\sigma$ has
been numerically calculated in \cite{Panos2}.

In figure \ref{fig:sigma10} we present the results for the cases
$\sigma=0.1, N=1$ and $\sigma=1, N=2$. The first case satisfies
$\sigma<2/N$ and the second satisfies $\sigma\geq 2/N$, in its
critical value $\sigma=\frac{2}{N}$. The threshold (\ref{newo}) is
fulfilled in both cases. We observe that in the case $\sigma=0.1,
N=1$, the threshold (\ref{newo}) is a sharp estimate of the power
observed for small values of frequency. These breathers serve this
time as real examples demonstrating that (\ref{newo}) predicts the
smallest value of a breather also in the case $\sigma<2/N$.  In the
second case we observe the appearance of the excitation threshold.

To show the global character of the estimate (\ref{newo}) as a
threshold on the existence of breather solutions, we consider in
Figure \ref{fig:sigma01} the ``limiting'' cases with respect to the
size of the nonlinearity exponent, $\sigma=0.1$ and $\sigma=10$, this
time for $N=2$. The first case is again an example for $\sigma<2/N$
and the second for $\sigma>2/N$. The threshold (\ref{newo}) is again
fulfilled in both cases. We observe that in both cases the threshold
(\ref{newo}) is still a quite sharp estimate of the smallest value of
the power of an existing breather as the comparison with the power of
breathers with small (large) values of frequency demonstrates.

It is worth noticing the existence of a maximum in the power for
$\sigma=1$, $N=2$, as was predicted in \cite{Panos3}.

\section*{Acknowledgements}  We would like to thank the referee for his valuable comments.  One of us (JC) acknowledges financial support from the MECD/FEDER
project FIS2004-01183.

\medskip

Received ; revised .

\medskip


\begin{thebibliography}{99}
\bibitem{RCarretal} (2230494) 
\newblock R. Carretero-Gonz\'{a}lez, J.D. Talley, C. Chong
  and B. A. Malomed, 
  \newblock \emph{Multistable solitons in the cubic-quintic
    discrete nonlinear Schr\"odinger Equation}, 
    \newblock Physica D, \textbf{216} (2006),
  77--89.

\bibitem{CJ} (0660633)
\newblock S. N. Chow and  J. K. Hale 
\newblock \emph{Methods of Bifurcation
    Theory}, 
    \newblock Grundlehren der mathematischen Wissenschaften -- A series of
  Comprehensive Studies in Mathematics {251}, Springer-Verlag, New-York,
  1982.

\bibitem{EilSatDNLS} 
\newblock J. Cuevas and J. C. Eilbeck, 
\newblock \emph{Discrete soliton
    collisions in a waveguide array with saturable nonlinearity},
  \newblock Phys. Lett. A, \textbf{358} (2006), 15--20.

\bibitem{DoriE} 
\newblock J. Dorignac, J. C. Eilbeck, M. Salerno and A.C. Scott,
  \emph{Quantum Signatures of Breather-Breather Interactions}, 
  \newblock Phys. Rev. Lett., \textbf{93} (2004), 025504.

\bibitem{Eil} 
\newblock J. C. Eilbeck and M. Johansson, 
\newblock \emph{The Discrete Nonlinear
    Schr\"odinger Equation-20 Years on}. `
    \newblock `Localization and Energy
  transfer in Nonlinear Systems'', eds L. V\'azquez, R.S. MacKay and M.P.
  Zorzano. World Scientific, Singapore, (2003) 44--67.

\bibitem{FlachMac} 
\newblock S. Flach, K. Kladko and R. S. MacKay, \emph{Energy
    thresholds for discrete breathers in one-, two-, and three
    dimensional lattices}, 
    \newblock Phys. Rev. Lett., \textbf{78} (1997), 1207--1210.

\bibitem{GiVel96} (1406282)
\newblock J. Ginibre and G. Velo, 
\newblock \emph{The Cauchy Problem in
    local spaces for the complex Ginzburg-Landau equation. I:
    Compactness methods}, 
    \newblock Physica D, \textbf{95} (1996), 191--228.

\bibitem{Maluckov1} 
\newblock L. Hadzievski, A. Maluckov, M. Stepic and D. Kip,
  \newblock \emph{Power controlled solitons stability and steering in lattices
    with saturable nonlinearity},
    \newblock Phys. Rev. Lett, \textbf{93} (2004), 033901.

\bibitem{speight} (1942954)
\newblock M. Haskins and J. M. Speight, 
\newblock \emph{Breather initial
    profiles in chains of weakly coupled anharmonic oscillators}, 
    \newblock Phys. Letters A, \textbf{299} (2002), 549--557.

\bibitem{K1} (2202146)
\newblock N. I. Karachalios, 
\newblock \emph{A remark on the existence of
    breather solutions for the Discrete Nonlinear Schr\"{o}dinger
    Equation: The case of site dependent anharmonic parameter}, 
    \newblock Proc.
  Edinburgh Math. Society, \textbf{49} (2006), 115--129.

\bibitem{Kevrekidis} 
\newblock P. G. Kevrekidis, K. \O. Rasmussen and A. R.
  Bishop, 
  \newblock \emph{The discrete nonlinear Schr\"odinger equation: A survey
    of recent results}, 
    \newblock Int. Journal of Modern Physics B, {\bf 15}
  (2001), 2833--2900
  
\bibitem{Panos3} 
\newblock P.G. Kevrekidis, K. \O. Rasmussen and A. R. Bishop,
  \emph{Two-dimensional discrete breathers: Construction, stability,
    and bifurcations}, 
    \newblock Phys. Rev. E, \textbf{61} (2000), 2006--2009.
    
\bibitem{Panos2} 
\newblock P.G. Kevrekidis, K.\ O. Rasmussen and A. R. Bishop,
  \newblock \emph{Localized excitations and their thresholds}, 
  \newblock Phys. Rev. E, \textbf{61} (2000), 4652--4655.
  
\bibitem{Maluckov3} (2230496)
\newblock A. Maluckov, L. Hadzievski and M. Stepic, 
\newblock \emph{Bifurcation analysis of the localized modes dynamics in lattices
    with saturable nonlinearity},
    \newblock Physica D, \textbf{216} (2006), 95--102.
    
\bibitem{Jes2} 
\newblock T. R. O. Melvin, A. R. Champneys, P. G. Kevrekidis and
  J.  Cuevas, 
  \newblock \emph{Radiationless travelling waves in saturable
    nonlinear Schr\"{o}dinger lattices},
    \newblock  Phys. Rev. Lett, \textbf{97} (2006),
  124101.
  
\bibitem{Mich} 
\newblock H. Michinel, J. Campo-T\'{a}boas, R.
  Garc\'{\i}a-Fern\'andez, J. R. Salgueiro and M. L. Quiroga-Teixeiro,
 \newblock \emph{Liquid light condensates}, 
 \newblock Phys. Rev. E, \textbf{65} (2002), 066604 .
 
\bibitem{ree79} (0751959)
\newblock M. Reed and B. Simon, 
\newblock \emph{Methods of Mathematical
    Physics I: Functional Analysis}, 
    \newblock Academic Press, New York, 1979.
    
\bibitem{st88} 
\newblock A. J. Sievers and S. Takeno, 
\newblock \emph{Intrinsic Localized
    Modes in anharmonic crystals}, 
    \newblock Phys. Rev. Lett., \textbf{61} (1988), 970--973.
    
\bibitem{Maluckov2} 
\newblock M. Stepic, D. Kip, L. Hadzievski and A. Maluckov,
\newblock  \emph{One-dimensional bright discrete solitons in media with
    saturable nonlinearity},
    \newblock Phys. Rev. E, \textbf{69} (2004), 066618.
    
\bibitem{RodMagnus} 
\newblock R. A. Vicencio and M. Johansson, 
\newblock \emph{Discrete
    soliton mobility in two dimensional waveguide arrays with
    saturable nonlinearity}, 
    \newblock Phys. Rev. E,  \textbf{73} (2006), 046602.
    
\bibitem{Wein99} (1690199)
\newblock M. Weinstein, 
\newblock \emph{Excitation thresholds for nonlinear
    localized modes on lattices}, 
    \newblock Nonlinearity, \textbf{12} (1999), 673--691.
    
\bibitem{zei85} (1033497) 
\newblock E. Zeidler, 
\newblock \emph{Nonlinear Functional Analysis and its
    Applications, Vol. II/A. Linear monotone
    Operators.}
    \newblock Springer-Verlag, Berlin, 1990.

%
\end{thebibliography}
\end{document}